\documentclass[10pt,aps,prl,twocolumn,notitlepage,showpacs,superscriptaddress]{revtex4-1}
\usepackage{amsthm,amsmath,amsfonts,amssymb,verbatim, color}
\usepackage{graphicx}
\usepackage{bm}
\usepackage{epsfig,slashed}
\usepackage[T1]{fontenc}
\usepackage[colorlinks=true,citecolor=blue,linkcolor=blue,urlcolor=blue]{hyperref}
\begin{document}

\newcommand{\tr}{\mathop{\mathrm{Tr}}}
\newcommand{\bsigma}{\boldsymbol{\sigma}}
\newcommand{\re}{\mathop{\mathrm{Re}}}
\newcommand{\im}{\mathop{\mathrm{Im}}}
\renewcommand{\b}[1]{{\boldsymbol{#1}}}
\newcommand{\diag}{\mathrm{diag}}
\newcommand{\sign}{\mathrm{sign}}
\newcommand{\sgn}{\mathop{\mathrm{sgn}}}
\renewcommand{\c}[1]{\mathcal{#1}}
\renewcommand{\d}{\text{\dj}}

\newcommand{\mb}{\bm}
\newcommand{\ua}{\uparrow}
\newcommand{\da}{\downarrow}
\newcommand{\ra}{\rightarrow}
\newcommand{\la}{\leftarrow}
\newcommand{\mc}{\mathcal}
\newcommand{\bs}{\boldsymbol}
\newcommand{\lra}{\leftrightarrow}
\newcommand{\nn}{\nonumber}
\newcommand{\half}{{\textstyle{\frac{1}{2}}}}
\newcommand{\mf}{\mathfrak}
\newcommand{\MF}{\text{MF}}
\newcommand{\IR}{\text{IR}}
\newcommand{\UV}{\text{UV}}

\DeclareGraphicsExtensions{.png}

\title{Landau theory of helical Fermi liquids}

\author{Rex Lundgren}
\affiliation{Department of Physics, The University of Texas at Austin, Austin, Texas 78712, USA}

\author{Joseph Maciejko}
\email[electronic address: ]{maciejko@ualberta.ca}
\affiliation{Department of Physics, University of Alberta, Edmonton, Alberta T6G 2E1, Canada}
\affiliation{Theoretical Physics Institute, University of Alberta, Edmonton, Alberta T6G 2E1, Canada}
\affiliation{Canadian Institute for Advanced Research, Toronto, Ontario M5G 1Z8, Canada}

\date\today

\begin{abstract}
Landau's phenomenological theory of Fermi liquids is a fundamental paradigm in many-body physics that has been remarkably successful in explaining the properties of a wide range of interacting fermion systems, such as liquid helium-3, nuclear matter, and electrons in metals. The $d$-dimensional boundaries of $(d+1)$-dimensional topological phases of matter such as quantum Hall systems and topological insulators provide new types of many-fermion systems that are topologically distinct from conventional $d$-dimensional many-fermion systems. We construct a phenomenological Landau theory for the two-dimensional helical Fermi liquid found on the surface of a three-dimensional time-reversal invariant topological insulator. In the presence of rotation symmetry, interactions between quasiparticles are described by ten independent Landau parameters per angular momentum channel, by contrast with the two (symmetric and antisymmetric) Landau parameters for a conventional spin-degenerate Fermi liquid. We then project quasiparticle states onto the Fermi surface and obtain an effectively spinless, projected Landau theory with a single projected Landau parameter per angular momentum channel that captures the spin-momentum locking or nontrivial Berry phase of the Fermi surface. As a result of this nontrivial Berry phase, projection to the Fermi surface can increase or lower the angular momentum of the quasiparticle interactions. We derive equilibrium properties, criteria for Fermi surface instabilities, and collective mode dispersions in terms of the projected Landau parameters. We briefly discuss experimental means of measuring projected Landau parameters.
\end{abstract}


\maketitle

The Landau theory of Fermi liquids (FL)~\cite{PinesNozieres}, or FL theory for short, is the cornerstone of our understanding of weakly correlated, gapless Fermi systems at low temperatures, such as $^3$He atoms in the normal liquid state and itinerant electrons in metals. FL theory explains the puzzling observation that despite strong interactions between the constituent fermions, many Fermi systems behave essentially as free Fermi gases, except for the renormalization of their physical properties which is captured by dimensionless quantities known as Landau parameters. These Landau parameters describe how the elementary excitations of the FL---the quasiparticles and quasiholes---interact with one another.

Topological insulators~\cite{hasan2010,*qi2011} provide new types of gapless Fermi systems: topological surface/edge states. In the absence of interparticle interactions, electrons propagating on the edge of a two-dimensional (2D) topological insulator~\cite{maciejko2011} form a 1D helical Fermi gas~\cite{kane2005}. In the presence of interactions, the 1D helical Fermi gas becomes a 1D helical Luttinger liquid~\cite{xu2006,*wu2006} with no sharply defined Fermi points. In 3D topological insulators, surface electrons form a 2D helical Fermi gas~\cite{fu2007}, which is expected to evolve adiabatically into a 2D helical FL in the presence of electron-electron interactions.

This paper presents a FL theory for the interacting 2D surface states of the 3D topological insulator. To our knowledge, such a helical FL theory has been missing in the literature despite the recent surge of interest in the effects of electron-electron interactions in topological insulators~\cite{InteractingTI}. In the spirit of standard FL theory~\cite{PinesNozieres}, we focus on systems with a discrete time-reversal symmetry, the protecting symmetry of topological insulators, as well as continuous translation and spatial rotation symmetries. We further consider the simplest case of a single surface Fermi surface---denoted simply as the Fermi surface in the following---which by rotation symmetry must be circular. This does not apply to certain topological insulators whose Fermi surface is strongly anisotropic, such as Bi$_2$Te$_3$ with 0.67\% Sn doping~\cite{chen2009} where there are large hexagonal warping effects due to the rhombohedral crystal structure of the bulk material~\cite{fu2009}. However, in several other topological insulators such as Bi$_2$Se$_3$~\cite{xia2009,*hsieh2009,*pan2011}, Bi$_2$Te$_2$Se~\cite{neupane2012,*neupane2013}, Sb$_x$Bi$_{2-x}$Se$_2$Te~\cite{neupane2012}, Bi$_{1.5}$Sb$_{0.5}$Te$_{1.7}$Se$_{1.3}$~\cite{kim2013}, Tl$_{1-x}$Bi$_{1+x}$Se$_{2-\delta}$~\cite{kuroda2013}, strained $\alpha$-Sn on InSb(001)~\cite{barfuss2013}, and strained HgTe~\cite{crauste2013}, the Fermi surface as observed in angle-resolved photoemission spectroscopy (ARPES) is very nearly circular. However, due to spin-momentum locking in the topological surface states~\cite{fu2007}---a consequence of strong spin-orbit coupling, rotation symmetry in a helical FL must necessarily involve spin degrees of freedom, which leads to a theory rather different from that of the conventional spin-degenerate FL. Moreover, the existence of a single nondegenerate Fermi surface---a consequence of the topological character of the bulk---eventually leads, via the application of the general principles of FL theory, to an {\it effectively spinless} FL theory. The physical properties of the resulting helical FL are nevertheless distinct from those of a truly spinless FL, due to a nontrivial mapping between physical, spinful quasiparticles, and the effective, spinless quasiparticles. For the same reason, our helical FL theory is also qualitatively different from recently constructed FL theories of non-topological spin-orbit coupled systems such as the Rashba 2D electron gas~\cite{ashrafi2013} and 3D spin-orbit coupled metals~\cite{fu2015}, which are characterized by two (spin-split) Fermi surfaces.

FL theory views the many-fermion system as a gas of elementary excitations above the ground state, the quasiparticles. Because translation symmetry is assumed, the momentum $\b{p}=(p_x,p_y)$ of the quasiparticles is well-defined and a configuration of quasiparticles is specified by a distribution function $n_\b{p}$. In a conventional FL, spin is conserved and the distribution function is diagonal in spin space $n_{\b{p}\sigma}=\langle c_{\b{p}\sigma}^\dag c_{\b{p}\sigma}^{\phantom\dagger}\rangle$, where $c_{\b{p}\sigma}^\dag$ ($c_{\b{p}\sigma}^{\phantom\dagger}$) is a creation (annihilation) operator for a fermion with momentum $\b{p}$ and spin $\sigma=\ua,\da$, but in systems with spin-orbit coupling such as the helical FL the distribution function is generally a matrix in spin space, $n_\b{p}^{\alpha\beta}=\langle c_{\b{p}\alpha}^\dag c_{\b{p}\beta}^{\phantom\dagger}\rangle$~\cite{fu2015}. The central quantity in FL theory is the energy $\delta E$ of the gas of interacting quasiparticles relative to the ground-state energy, expressed as a functional of the deviation $\delta n_\b{p}^{\alpha\beta}\equiv n_\b{p}^{\alpha\beta}-n_\b{p}^{(0)\alpha\beta}$ of the distribution function from its value in the ground state,
\begin{align}\label{dE_unprojected}
\delta E[\delta n_\b{p}]=&\int\d p\,h_{\alpha\beta}(\b{p})\delta n_\b{p}^{\alpha\beta}\nonumber\\
&+\half\int\d p\,\d p'\,V_{\alpha\beta;\gamma\delta}(\hat{\b{p}},\hat{\b{p}}')\delta n_\b{p}^{\alpha\beta}\delta n_{\b{p}'}^{\gamma\delta},
\end{align}
where (working in units such that $\hbar=1$)
\begin{align}\label{HDirac}
h(\b{p})=v_F\hat{\b{z}}\cdot(\bsigma\times\b{p}),
\end{align}
is the single-particle Dirac Hamiltonian of the topological surface state~\cite{hasan2010,*qi2011} with $v_F$ the Fermi velocity~\cite{RenFermiVelNote}, $V_{\alpha\beta;\gamma\delta}(\hat{\b{p}},\hat{\b{p}}')$ is a reduced two-body interaction that depends only on the unit vector $\hat{\b{p}}\equiv\b{p}/|\b{p}|$ parameterizing the Fermi surface, and we denote the integration measure by $\int\d p\equiv\int\frac{d^2p}{(2\pi)^2}$. The form of Eq.~(\ref{dE_unprojected}) can be obtained from a generic, translationally invariant interaction $V_{\alpha\beta;\gamma\delta}(\b{k},\b{k}',\b{q})$ by requiring that all fermionic momenta lie on the Fermi surface~\cite{SuppMat}.

Our first goal is to derive the most general form of the two-body interaction $V_{\alpha\beta;\gamma\delta}(\hat{\b{p}},\hat{\b{p}}')$ consistent with the general principles of quantum mechanics and the symmetries of the problem. This goal is most easily achieved by expanding the two-body interaction as
\begin{align}\label{VExpand}
V_{\alpha\beta;\gamma\delta}(\hat{\b{p}},\hat{\b{p}}')=\sum_{\mu,\nu=0}^3\sum_{l,l'=-\infty}^\infty V_{\mu\nu}^{ll'}e^{i(l\theta_\b{p}+l'\theta_{\b{p}'})}\sigma_{\alpha\beta}^\mu\sigma_{\gamma\delta}^\nu,
\end{align}
where $\hat{\b{p}}=(\cos\theta_\b{p},\sin\theta_\b{p})$, $l,l'$ are angular momentum quantum numbers, and the set of four $2\times 2$ Hermitian matrices $\sigma^\mu=(1,\bsigma)$ where $1$ denotes the identity matrix allows us to construct the quasiparticle charge $\delta\rho_\b{p}$ and spin $\delta s^i_\b{p}$ densities ($i=x,y,z$),
\begin{align}\label{DefRhoS}
\delta\rho_\b{p}=\sigma_{\alpha\beta}^0\delta n_\b{p}^{\alpha\beta}=\delta_{\alpha\beta}\delta n_\b{p}^{\alpha\beta},\hspace{5mm}
\delta s^i_\b{p}=\half\sigma_{\alpha\beta}^i\delta n_\b{p}^{\alpha\beta}.
\end{align}
Upon substituting Eq.~(\ref{VExpand}) in Eq.~(\ref{dE_unprojected}), one obtains three classes of terms: charge-charge interactions proportional to $V_{00}^{ll'}$, spin-spin interactions proportional to $V_{ij}^{ll'}$, and spin-charge interactions proportional to $V_{0i}^{ll'}=V_{i0}^{l'l}$. Time-reversal symmetry implies that the angular momenta $l$ and $l'$ must differ by an even integer for charge-charge and spin-spin interactions and by an odd integer for spin-charge interactions~\cite{SuppMat}.

The main difference between a conventional FL and a spin-orbit coupled FL such as the helical FL lies in the consequences of rotation symmetry. The single-particle Hamiltonian (\ref{HDirac}) is neither invariant under a spatial rotation nor under a spin rotation, but is invariant under a simultaneous rotation of spatial and spin coordinates: $[J_z,h(\b{p})]=0$, where $J_z=-i\frac{\partial}{\partial\theta_\b{p}}+\half\sigma^z$ is the total (orbital plus spin) angular momentum in the $z$ direction. Requiring that the interaction term in Eq.~(\ref{dE_unprojected}) be also invariant under such rotations, we find that it can be written as the sum of three terms $\delta V_{cc}$, $\delta V_{sc}$, and $\delta V_{ss}$, where~\cite{SuppMat}
\begin{align}\label{Vcc}
\delta V_{cc}=\half\sum_{l=0}^\infty\int\d p\,\d p'\,f^{cc}_l\cos l\theta_{\b{p}\b{p}'}\delta\rho_\b{p}\delta\rho_{\b{p}'},
\end{align}
is the charge-charge interaction,
\begin{align}\label{Vsc}
\delta &V_{sc}=\sum_{l=0}^\infty\int\d p\,\d p'\nonumber\\
&\times\Bigl[(f^{sc,1}_l\cos l\theta_{\b{p}\b{p}'}+f^{sc,2}_l\sin l\theta_{\b{p}\b{p}'})\delta\rho_\b{p}\hat{\b{p}}'\cdot\delta\b{s}_{\b{p}'}\nonumber\\
&+(f^{sc,3}_l\cos l\theta_{\b{p}\b{p}'}+f^{sc,4}_l\sin l\theta_{\b{p}\b{p}'})\delta\rho_\b{p}\hat{\b{p}}'\times\delta\b{s}_{\b{p}'}\Bigr],
\end{align}
is the spin-charge interaction, and
\begin{widetext}
\begin{align}\label{Vss}
&\delta V_{ss}=\half\sum_{l=0}^\infty\int\d p\,\d p'
\Bigl\{\cos l\theta_{\b{p}\b{p}'}\Bigl(f^{ss,1}_l(\delta s_\b{p}^x\delta s_{\b{p}'}^x+\delta s_\b{p}^y\delta s_{\b{p}'}^y)
+f^{ss,2}_l\delta s^z_\b{p}\delta s^z_{\b{p}'}\Bigr)+f^{ss,3}_l\sin l\theta_{\b{p}\b{p}'}\delta\b{s}_\b{p}\times\delta\b{s}_{\b{p}'}\nonumber\\
&+\cos l\theta_{\b{p}\b{p}'}\Bigl(f^{ss,4}_l\left[\left(\hat{\b{p}}\cdot\delta\b{s}_\b{p}\right)
\left(\hat{\b{p}}'\times\delta\b{s}_{\b{p}'}\right)+\left(\hat{\b{p}}\times\delta\b{s}_\b{p}\right)
\left(\hat{\b{p}}'\cdot\delta\b{s}_{\b{p}'}\right)\right]
+f^{ss,5}_l\left[\left(\hat{\b{p}}\cdot\delta\b{s}_\b{p}\right)
\left(\hat{\b{p}}'\cdot\delta\b{s}_{\b{p}'}\right)-\left(\hat{\b{p}}\times\delta\b{s}_\b{p}\right)
\left(\hat{\b{p}}'\times\delta\b{s}_{\b{p}'}\right)\right]\Bigr)\Bigr\},
\end{align}
\end{widetext}
is the spin-spin interaction. We denote by $\theta_{\b{p}\b{p}'}\equiv\theta_{\b{p}'}-\theta_\b{p}$ the relative angle between $\hat{\b{p}}$ and $\hat{\b{p}}'$, and write $\b{a}\times\b{b}\equiv\hat{\b{z}}\cdot(\b{a}\times\b{b})$ for the cross product of two in-plane vectors.

Equations~(\ref{Vcc})-(\ref{Vss}), the first main result of this work, represent the most general short-range two-body interaction in a helical FL consistent with translation, rotation, and time-reversal symmetries. The interaction is specified by ten real Landau parameters for each value of the relative angular momentum $l=0,1,2,\ldots$: one charge-charge parameter $f^{cc}_l$, four spin-charge parameters $f^{sc,1}_l,\ldots,f^{sc,4}_l$, and five spin-spin parameters $f^{ss,1}_l,\ldots,f^{ss,5}_l$. This stands in contrast to the two Landau parameters $f_l^s$ (spin symmetric) and $f_l^a$ (spin antisymmetric) in a conventional FL~\cite{PinesNozieres}, which would correspond to $f_l^s=f_l^{cc}$ and $f_l^a=\frac{1}{4}f_l^{ss,1}=\frac{1}{4}f_l^{ss,2}$ in the absence of spin-orbit coupling. In particular, spin-orbit coupling allows for a nonzero spin-charge interaction (\ref{Vsc}) which would be forbidden by separate spatial and spin rotation symmetries in a conventional FL. The spin-spin interaction (\ref{Vss}) also exhibits novel features: $f_l^{ss,1}\neq f_l^{ss,2}$ in general, which corresponds to an XXZ interaction with Ising anisotropy rather than the conventional $SU(2)$-symmetric Heisenberg interaction; $f_l^{ss,3}$ is a Dzyaloshinskii-Moriya interaction; and $f_l^{ss,4}$, $f_l^{ss,5}$ are anisotropic spin-spin interactions similar to those found in compass models~\cite{nussinov2015}, but with a continuous rather than discrete spin-orbit rotation symmetry.

While Eq.~(\ref{Vcc})-(\ref{Vss}) in conjunction with Eq.~(\ref{dE_unprojected}) correctly describe the helical FL, in the spirit of FL theory one can go one step further and only retain electron states on the Fermi surface. Because of the strong spin-orbit coupling present in the Dirac Hamiltonian (\ref{HDirac}), such electrons are annihilated by the operator $\psi_{\b{p}\pm}=\frac{1}{\sqrt{2}}(ie^{-i\theta_\b{p}}c_{\b{p}\ua}\pm c_{\b{p}\da})$, where positive ($+$) helicity corresponds to a positive Fermi energy $\epsilon_F>0$ above the Dirac point, and negative ($-$) helicity corresponds to a negative Fermi energy $\epsilon_F<0$. Inverting this relation, one can express the spin eigenoperators $c_{\b{p}\sigma}$ in terms of the helicity eigenoperators $\psi_{\b{p}\pm}$ as $c_{\b{p}\ua}=\frac{ie^{-i\theta_\b{p}}}{\sqrt{2}}(\psi_{\b{p}+}+\psi_{\b{p}-})$ and $c_{\b{p}\da}=\frac{1}{\sqrt{2}}(\psi_{\b{p}+}-\psi_{\b{p}-})$. Choosing $\epsilon_F>0$ for definiteness, the Fermi surface consists exclusively of electron states of positive helicity, such that one may wish to drop the negative helicity eigenoperators $\psi_{\b{p}-}$ entirely from these expressions for $c_{\b{p}\ua}$ and $c_{\b{p}\da}$. Applying this procedure to Eq.~(\ref{dE_unprojected}) yields a Landau functional for an effectively spinless FL theory,
\begin{align}\label{ProjFL}
\delta\bar{E}[\delta\bar{n}_\b{p}]=&\int\d p\,\epsilon_\b{p}^0\delta\bar{n}_\b{p}\nonumber\\
&+\half\sum_{l=0}^\infty\int\d p\,\d p'\bar{f}_l\cos l\theta_{\b{p}\b{p}'}\delta\bar{n}_\b{p}\delta\bar{n}_{\b{p}'},
\end{align}
where $\epsilon_\b{p}^0=v_F|\b{p}|$ is the dispersion relation of positive helicity quasiparticles, $\delta\bar{n}_\b{p}=\bar{n}_\b{p}-\bar{n}_\b{p}^{(0)}$ with $\bar{n}_\b{p}\equiv\langle\psi_{\b{p}+}^\dag\psi_{\b{p}+}^{\phantom\dagger}\rangle$ is the distribution function for these quasiparticles, and $\bar{f}_l$ are effectively spinless, projected Landau parameters related to the ten unprojected Landau parameters previously discussed by
\begin{align}\label{ProjLandParam}
\bar{f}_l&=f_l^{cc}-f_l^{sc,3}-{\textstyle{\frac{1}{4}}}f_l^{ss,5}\nonumber\\
&\hspace{5mm}+{\textstyle{\frac{1}{8}}}(f_{l-1}^{ss,1}-f_{l-1}^{ss,3}+f_{l+1}^{ss,1}+f_{l+1}^{ss,3}),
\end{align}
for $l=0,1,2,\ldots$, with the definition $f_{-1}^{ss,1}=f_{-1}^{ss,3}\equiv 0$. The quasiparticle charge and spin densities (\ref{DefRhoS}) are given in terms of $\delta\bar{n}_\b{p}$ by
\begin{align}\label{RhoSProj}
\delta\rho_\b{p}=\delta\bar{n}_\b{p},\hspace{3mm}
\delta s_\b{p}^i=\half\epsilon_{ij}\hat{p}_j\delta\bar{n}_\b{p},\,i=x,y,\hspace{3mm}
\delta s_\b{p}^z=0,
\end{align}
where the last two equalities express spin-momentum locking in the $xy$ plane. Equations (\ref{ProjFL})-(\ref{RhoSProj}), together with the definitions of the unprojected Landau parameters in Eq.~(\ref{Vcc})-(\ref{Vss}), are the second main result of this work.

Before deriving the physical properties of the helical FL from the projected Landau functional (\ref{ProjFL}), we pause to discuss a number of interesting features of the relationship (\ref{ProjLandParam}) between projected and unprojected Landau parameters. The unprojected Landau parameters $f_l^{sc,1}$, $f_l^{sc,2}$, and $f_l^{ss,4}$ do not enter the projected interaction because spin and momentum are perpendicular on the Fermi surface ($\hat{\b{p}}\cdot\delta\b{s}_\b{p}=0$) due to spin-momentum locking. The parameter $f_l^{sc,4}$ does not enter either because it produces a projected interaction that is odd under $\b{p}\leftrightarrow\b{p}'$, which is inconsistent with particle indistinguishability. The last term on the right-hand side of Eq.~(\ref{ProjLandParam}) shows that projection to the Fermi surface can effectively raise or lower the angular momentum of the unprojected interaction. For example, for $l=1$ one has
\begin{align}
\bar{f}_1&=f_1^{cc}-f_1^{sc,3}-{\textstyle{\frac{1}{4}}}f_1^{ss,5}\nonumber\\
&\hspace{5mm}+{\textstyle{\frac{1}{8}}}(f_0^{ss,1}-f_0^{ss,3}+f_{2}^{ss,1}+f_{2}^{ss,3}),
\end{align}
that is, an isotropic, $s$-wave ($l=0$) microscopic interaction can produce an anisotropic, $p$-wave ($l=1$) effective interaction in the projected theory. This can be seen as the particle-hole counterpart to the effective $p$-wave interaction in the Bardeen-Cooper-Schrieffer (BCS) channel produced on the doped surface of a 3D topological insulator by a microscopic $s$-wave BCS interaction~\cite{fu2008}.

As in standard FL theory, many physical properties of the helical FL can be derived from the projected Landau functional (\ref{ProjFL}). The simplest property is Luttinger's theorem~\cite{luttinger1960}, i.e., the relation $p_F=\sqrt{4\pi n}$ between Fermi momentum $p_F$ and total density $n$ of quasiparticles, which is also equal to the total density of electrons (defining a system with $p_F=0$ as the vacuum). That Luttinger's theorem holds in its original form despite the presence of strong spin-orbit coupling is a consequence of the existence of a single helical Fermi surface, which is only possible on the surface of a 3D topological phase. Interactions in topologically trivial spin-orbit coupled systems such as the Rashba 2D electron gas can individually renormalize the Fermi momenta of the two spin-split Fermi surfaces~\cite{ashrafi2013}. Other equililibrium properties of the helical FL can be calculated from the quasiparticle energy $\epsilon_\b{p}$, defined as the functional derivative of the Landau functional with respect to the distribution function,
\begin{align}\label{RenEp}
\epsilon_\b{p}=\frac{\delta\bar{E}}{\delta\bar{n}_\b{p}}=\epsilon_\b{p}^0+\sum_{l=0}^\infty\int\d p'\bar{f}_l\cos l\theta_{\b{p}\b{p}'}\delta\bar{n}_{\b{p}'}.
\end{align}
From Eq.~(\ref{RenEp}) one can follow the standard FL approach~\cite{SuppMat} to derive the electronic specific heat coefficient $\gamma\equiv c_v/T$ and electronic compressibility $\kappa$ of the helical FL at zero temperature,
\begin{align}\label{SpecHeatComp}
\gamma={\textstyle{\frac{1}{3}}}\pi^2k_B^2 \rho(\epsilon_F),\hspace{5mm}
\kappa=\frac{\rho(\epsilon_F)}{n^2}\frac{1}{1+\bar{F}_0},
\end{align}
where we define dimensionless Landau parameters $\bar{F}_0\equiv \rho(\epsilon_F)\bar{f}_0$ and $\bar{F}_l\equiv\half \rho(\epsilon_F)\bar{f}_l$, $l=1,2,3,\ldots$, with $\rho(\epsilon_F)=\epsilon_F/2\pi v_F^2$ the density of states of the helical FL at the Fermi energy $\epsilon_F=v_Fp_F$. The compressibility becomes negative for $\bar{F}_0<-1$, signaling an instability towards phase separation~\cite{castro-neto1995}. Unlike in a standard FL, here this condition can be reached not only for attractive density-density interactions, but also as a result of spin-charge or even purely spin-spin interactions, given the relation (\ref{ProjLandParam}) between the projected and unprojected Landau parameters.

The renormalized Fermi velocity $v_F$ differs in general from the Fermi velocity of noninteracting electrons $v_F^0$. This is similar in spirit to the renormalization of the quasiparticle mass in a standard FL. The derivation of the latter relies on Galilean invariance, while in the helical FL, Galilean invariance is broken by spin-orbit coupling. However, adiabatic continuity still implies that the total flux of quasiparticles is equal to the total flux of electrons~\cite{PinesNozieres}. The latter is calculated from the quantum-mechanical velocity operator for electrons $\b{v}_e=v_F^0(\hat{\b{z}}\times\bsigma)$ which, for momentum-independent microscopic interactions~\cite{MomIndInt}, is the same as in the absence of interactions~\cite{raghu2010}: it is a function of the noninteracting Fermi velocity, rather than the renormalized one. The total quasiparticle flux is a function of the quasiparticle velocity $\b{v}_\textrm{qp}=\nabla_\b{p}\epsilon_\b{p}$. Equating the two fluxes yields a relation between the two Fermi velocities~\cite{SuppMat},
\begin{align}\label{vFren}
\frac{v_F^0}{v_F}=1+\bar{F}_1,
\end{align}
which is the helical FL analog of the relation $\frac{m^*}{m}=1+\frac{1}{3}F_1^s$ between renormalized $m^*$ and noninteracting $m$ quasiparticle masses in a standard FL~\cite{PinesNozieres}.

The spin susceptibility introduces some added subtleties: unlike in a standard FL, it is not, strictly speaking, a Fermi surface property. Indeed, it depends explicitly on a high-energy cutoff $\Lambda$ already in the noninteracting limit~\cite{vazifeh2012,zhao2014}. In a standard FL, one can always choose the spin quantization axis to be parallel to the applied magnetic field $\b{B}$, such that the quasiparticle energy shift $\delta\epsilon_{\b{p}\sigma}=\frac{1}{2}g\mu_B B\sigma$ due to Zeeman coupling ($g$ is the $g$-factor, $\mu_B$ is the Bohr magneton) is diagonal in the spin basis $\sigma=\pm 1$. The resulting change in occupation numbers is localized to the Fermi surface in the zero-field limit, causing the spin susceptibility to be a Fermi surface property. In the helical FL, there is no freedom to choose the spin quantization axis due to spin-momentum locking, and the Zeeman coupling contains off-diagonal terms in the helicity basis. The projected FL theory (\ref{ProjFL}), which projects out negative helicity states, cannot take these off-diagonal terms into account and thus should not be expected to yield exact results for the spin susceptibility. Nevertheless, one can calculate the Fermi surface contribution to the spin susceptibility using (\ref{ProjFL}) and compare it in the noninteracting limit to an exact calculation that takes both helicities into account. The spin susceptibility tensor $\chi_{ij}$ is found to be diagonal, with in-plane $\chi_{xx}=\chi_{yy}$ and out-of-plane $\chi_{zz}$ components given by
\begin{align}\label{ChiProj}
\chi_{xx}={\textstyle{\frac{1}{8}}}g^2\mu_B^2 \rho(\epsilon_F)\frac{1}{1+\bar{F}_1},\hspace{5mm}
\chi_{zz}=0,
\end{align}
in the projected FL theory, and
\begin{align}\label{ChiExact}
\chi_{xx}={\textstyle{\frac{1}{8}}}g^2\mu_B^2 \rho(\Lambda),\hspace{5mm}
\chi_{zz}={\textstyle{\frac{1}{4}}}g^2\mu_B^2[\rho(\Lambda)-\rho(\epsilon_F)],
\end{align}
for the noninteracting Dirac surface state, including both helicities~\cite{SuppMat}. Thus in the noninteracting limit, Eq.~(\ref{ChiProj}) and (\ref{ChiExact}) agree in the formal limit of large Fermi energy $\epsilon_F\rightarrow\Lambda$. By contrast with the spin susceptibility of the standard FL which is renormalized by the spin-antisymmetric $l=0$ Landau parameter $F_0^a$, here it is renormalized by a $l=1$ Landau parameter due to spin-momentum locking on the Fermi surface.

Pomeranchuk instabilities~\cite{pomeranchuk1958} are instabilities of the Fermi surface towards spontaneous, static distortions of its shape. To study such instabilities in the helical FL, one characterizes distortions of the Fermi surface by an angle-dependent Fermi momentum, expanded in angular momentum components,
\begin{align}
p_F(\theta)-p_F=\sum_{l=-\infty}^\infty A_l e^{il\theta},
\end{align}
where $A_{-l}=A_l^*$ because $p_F(\theta)$ is real. Substituting this expression into the Landau functional (\ref{ProjFL}), one finds that the energy $\delta\bar{E}$ remains positive, and thus the helical FL stable, if and only if~\cite{SuppMat}
\begin{align}
\bar{F}_l>-1,
\end{align}
for all $l=0,1,2,\ldots$ This is the same as Pomeranchuk's original criterion in 2D, but applied this time to the projected Landau parameters, which are nontrivial functions of the unprojected ones. It contains as special cases the instability towards phase separation, already seen, as well as an instability towards in-plane magnetic order~\cite{xu2010} for $\bar{F}_1\rightarrow -1$, that is signaled by divergences of the in-plane spin susceptibility (\ref{ChiProj}) and the renormalized Fermi velocity (\ref{vFren}). The latter divergence also accompanies the $l=1$ spin-symmetric Pomeranchuk instability of the standard FL~\cite{varma2005}. 
The $l=2$ instability is towards quadrupolar distortions of the helical Fermi surface, characterized in the projected FL theory by a nonzero value of the traceless, symmetric nematic order parameter $\bar{Q}_{ij}=\int\d p\,\bar{Q}_{ij}(\b{p})$ where $\bar{Q}_{ij}(\b{p})=(2\hat{p}_i\hat{p}_j-\delta_{ij})\delta\bar{n}_\b{p}$. This effectively spinless order parameter is identical to the one that describes nematic order in a standard spin-degenerate FL~\cite{oganesyan2001}. In the original unprojected theory however, this translates into a nonzero value of $Q_{ij}=\int\d p\,Q_{ij}(\b{p})$ where
\begin{align}
Q_{ij}(\b{p})=\hat{p}_i\delta s^j_\b{p}+\hat{p}_j\delta s^i_\b{p}-\delta_{ij}\hat{\b{p}}\cdot\delta\b{s}_\b{p},
\end{align}
is a quadrupolar order parameter involving both spatial and spin degrees of freedom that was recently discussed in the context of possible instabilities of surface Majorana fermions in the topological superfluid $^3$He-$B$~\cite{park2015} and 3D spin-orbit coupled metals~\cite{fu2015,norman2015}. Thus the quadrupolar distortion of a helical Fermi surface is necessarily accompanied by a time-reversal invariant form of magnetic order similar in spirit to spin nematic order~\cite{podolsky2005}.

Nonequilibrium properties of the helical FL such as collective modes can also be studied using the projected FL theory, assuming that the relaxation-time approximation is valid such that scattering between states of different helicities can be neglected. In the hydrodynamic regime $\omega\tau\ll 1$ where $\tau$ is the quasiparticle collision time, the helical FL supports ordinary sound waves (first sound) with velocity~\cite{SuppMat}
\begin{align}\label{c1}
c_1=v_F\sqrt{\half(1+\bar{F}_0)(1+\bar{F}_1)},
\end{align}
while in the collisionless regime $\omega\tau\gg 1$ a zero sound mode may exist under certain conditions~\cite{raghu2010}. If $\bar{F}_0>0$ only is nonzero, the zero sound velocity is given in the limits of strong and weak interactions by~\cite{SuppMat}
\begin{align}
c_0&\approx v_F\sqrt{\half\bar{F}_0},\hspace{5mm}\bar{F}_0\rightarrow\infty,\label{c0a}\\
c_0&\approx v_F\left(1+\half\bar{F}_0^2\right),\hspace{5mm}\bar{F}_0\rightarrow 0.\label{c0b}
\end{align}

We conclude by discussing prospects for the experimental determination of the projected Landau parameters $\bar{F}_l$. ARPES can determine $p_F$ which, via Luttinger's theorem, yields the density $n$. Using Eq.~(\ref{SpecHeatComp}), $\bar{F}_0$ could then be inferred from measurements of the heat capacity and electronic compressibility of the surface states. The latter can in principle be determined directly from the ARPES data or via single electron transistor microscopy~\cite{Abergel2012}. To determine $\bar{F}_1$, one could perform a transient spin grating experiment~\cite{raghu2010} to generate a spin-density wave with momentum $\b{q}$ and transverse amplitude $s_\b{q}^T$. Due to spin-momentum locking, this will induce a density wave at the same momentum with amplitude $n_\b{q}$. The existence of an undamped sound mode at frequency $\omega=c_sq$ implies a relation between the two amplitudes~\cite{SuppMat},
\begin{align}
\frac{s_\b{q}^T}{n_\b{q}}=\frac{1}{1+\bar{F}_1}\frac{c_s}{v_F},
\end{align}
where $c_s$ is either $c_1$ or $c_0$ depending on whether one is in the hydrodynamic or collisionless regime. Using Eq.~(\ref{c1})-(\ref{c0b}) one can extract $\bar{F}_1$ from a measurement of the amplitude ratio $s_\b{q}^T/n_\b{q}$ and previous knowledge of $\bar{F}_0$.

We acknowledge J. P. Davis, G. A. Fiete, and S. A. Kivelson for useful discussions. R.L. was supported by National Science Foundation (NSF) Graduate Research Fellowship award number 2012115499 and NSF Grant No.~DMR-0955778. J.M. was supported by the Natural Sciences and Engineering Research Council of Canada (NSERC), the Canada Research Chair Program (CRC), the Canadian Institute for Advanced Research (CIFAR), and the University of Alberta. This research was supported in part by Perimeter Institute for Theoretical Physics. Research at Perimeter Institute is supported by the Government of Canada through Industry Canada and by the Province of Ontario through the Ministry of Economic Development \& Innovation.

\bibliography{HFL}

\begin{thebibliography}{41}
\bibitem [{Ren()}]{RenFermiVelNote}%
  \BibitemOpen
  \href@noop {} {}\bibinfo {note} {Our basic assumption is that the
  quasiparticles of the \emph{interacting} Fermi liquid retain the helical spin
  texture implied by Eq.~(\ref{HDirac}), as is observed
  experimentally.}\BibitemShut {Stop}%
\bibitem [{Mom()}]{MomIndInt}%
  \BibitemOpen
  \href@noop {} {}\bibinfo {note} {The necessary condition is that the
  microscopic interaction Hamiltonian---the Coulomb interaction---only depends
  on the positions of the electrons and not their momenta. However, the Landau
  functional (\ref{dE_unprojected}), which could in principle be explicitly
  derived from the microscopic Hamiltonian using diagrammatic perturbation
  theory~\cite{AGD}, will in general contain momentum-dependent interaction
  terms [see Eq.~(\ref{Vcc})-(\ref{Vss})].}\BibitemShut {Stop}%
\end{thebibliography}

\end{document}


\renewcommand{\Re}{\mathop{\mathrm{Re}}}
\renewcommand{\Im}{\mathop{\mathrm{Im}}}
\renewcommand{\b}[1]{{\boldsymbol{#1}}}
\newcommand{\ua}{\uparrow}
\newcommand{\da}{\downarrow}
\renewcommand{\d}{\text{\dj}}
\newcommand{\bsigma}{\boldsymbol{\sigma}}
\newcommand{\blambda}{\boldsymbol{\lambda}}
\newcommand{\tr}{\mathop{\mathrm{Tr}}}
\newcommand{\sgn}{\mathop{\mathrm{sgn}}}
\newcommand{\sech}{\mathop{\mathrm{sech}}}
\newcommand{\diag}{\mathop{\mathrm{diag}}}
\newcommand{\half}{{\textstyle\frac{1}{2}}}
\newcommand{\sh}{{\textstyle{\frac{1}{2}}}}
\newcommand{\ish}{{\textstyle{\frac{i}{2}}}}
\newcommand{\thf}{{\textstyle{\frac{3}{2}}}}
\newcommand{\be}{\begin{equation}}
\newcommand{\ee}{\end{equation}}

\renewcommand{\thetable}{S\Roman{table}}
\renewcommand{\thefigure}{S\arabic{figure}}
\renewcommand{\thesection}{S\Roman{section}}
\renewcommand{\thesubsection}{S\arabic{subsection}}
\renewcommand{\theequation}{S\arabic{equation}}

\title{Supplemental material for ``Landau theory of helical Fermi liquids''}

\author{Rex Lundgren}
\affiliation{Department of Physics, The University of Texas at Austin, Austin, Texas 78712, USA}

\author{Joseph Maciejko}
\email[electronic address: ]{maciejko@ualberta.ca}
\affiliation{Department of Physics, University of Alberta, Edmonton, Alberta T6G 2E1, Canada}
\affiliation{Theoretical Physics Institute, University of Alberta, Edmonton, Alberta T6G 2E1, Canada}
\affiliation{Canadian Institute for Advanced Research, Toronto, Ontario M5G 1Z8, Canada}

\date\today

\maketitle

This supplemental material provides a detailed derivation of the results presented in the main text. In Sec.~\ref{sec:red2body}, we justify the form of the quasiparticle interaction term used in Eq.~(1) of the main text. In Sec.~\ref{sec:constrainsymm} and Sec.~\ref{sec:landau}, we constrain the form of the interaction term by symmetries and derive Eq.~(5)-(9) of the main text. In Sec.~\ref{sec:eqprop}, we use the projected Landau functional to derive the equilibrium properties of the helical Fermi liquid, i.e., Eq.~(13)-(15), (18), and (19) of the main text. In Sec.~\ref{sec:spinfreedirac} we study the spin susceptibility of the noninteracting Dirac cone and derive Eq.~(16) of the main text. Finally, in Sec.~\ref{Coll_MODES} we study the collective modes (sound modes) of the helical Fermi liquid and derive Eq.~(20)-(23) of the main text.

\section{Reduced two-body interaction}
\label{sec:red2body}

In this section we derive the reduced two-body interaction [Eq.~(1) in the main text] from a generic, translationally invariant two-body interaction. In this supplemental material we use the language of second-quantized interaction Hamiltonians, but the same reasoning applies to the interaction term in the Landau functional. A generic translationally invariant interaction is given by
\begin{align}\label{genericV}
V=\half\int\frac{d^2p}{(2\pi)^2}\int\frac{d^2p'}{(2\pi)^2}\int\frac{d^2q}{(2\pi)^2}V_{\alpha\beta;\gamma\delta}(\b{p},\b{p}',\b{q})c_{\b{p}+\b{q},\alpha}^\dag c_{\b{p}\beta}^{\phantom\dagger}c_{\b{p}'-\b{q},\gamma}^\dag c_{\b{p}'\delta}^{\phantom\dagger}.
\end{align}
In Fermi liquid theory, one considers the low-temperature limit where all fermionic momenta must lie on the Fermi surface,\cite{PinesNozieres} which implies that there are only three possible interaction channels: forward scattering with $\b{q}=0$, exchange scattering with $\b{q}=\b{p}'-\b{p}$, and the Bardeen-Cooper-Schrieffer (BCS) channel with $\b{p}'=-\b{p}$ but $\b{q}$ otherwise arbitrary. Since we are not interested in pairing and superconductivity in the present work, we forget about the BCS channel. Therefore at low energies we can simplify the interaction to include only forward and exchange scattering,
\begin{align}\label{Vapprox}
V&\approx\half\int\frac{d^2p}{(2\pi)^2}\int\frac{d^2p'}{(2\pi)^2}\left(V_{\alpha\beta;\gamma\delta}(\b{p},\b{p}',0)c_{\b{p}\alpha}^\dag c_{\b{p}\beta}^{\phantom\dagger}c_{\b{p}'\gamma}^\dag c_{\b{p}'\delta}^{\phantom\dagger}+V_{\alpha\beta;\gamma\delta}(\b{p},\b{p}',\b{p}'-\b{p})c_{\b{p}'\alpha}^\dag c_{\b{p}\beta}^{\phantom\dagger}c_{\b{p}\gamma}^\dag c_{\b{p}'\delta}^{\phantom\dagger}\right)\nonumber\\
&=\half\int\frac{d^2p}{(2\pi)^2}\int\frac{d^2p'}{(2\pi)^2}\left(V_{\alpha\beta;\gamma\delta}(\b{p},\b{p}',0)c_{\b{p}\alpha}^\dag c_{\b{p}\beta}^{\phantom\dagger}c_{\b{p}'\gamma}^\dag c_{\b{p}'\delta}^{\phantom\dagger}-V_{\alpha\beta;\gamma\delta}(\b{p},\b{p}',\b{p}'-\b{p})c_{\b{p}\gamma}^\dag c_{\b{p}\beta}^{\phantom\dagger}c_{\b{p}'\alpha}^\dag  c_{\b{p}'\delta}^{\phantom\dagger}\right),
\end{align}
where we have ignored one-body terms. One can show from Eq.~(\ref{genericV}) that Fermi statistics implies
\begin{align}
V_{\alpha\beta;\gamma\delta}(\b{p},\b{p}',\b{q})=-V_{\gamma\beta;\alpha\delta}(\b{p},\b{p}',\b{p}'-\b{p}-\b{q}),
\end{align}
which upon setting $\b{q}=0$ yields $V_{\alpha\beta;\gamma\delta}(\b{p},\b{p}',0)=-V_{\gamma\beta;\alpha\delta}(\b{p},\b{p}',\b{p}'-\b{p})$, i.e., the forward scattering and exchange scattering contributions are related. Substituting this into Eq.~(\ref{Vapprox}), we find that the two contributions are in fact equal and simply add,
\begin{align}
V=\int\frac{d^2p}{(2\pi)^2}\int\frac{d^2p'}{(2\pi)^2}V_{\alpha\beta;\gamma\delta}(\b{p},\b{p}',0)c_{\b{p}\alpha}^\dag c_{\b{p}\beta}^{\phantom\dagger}c_{\b{p}'\gamma}^\dag c_{\b{p}'\delta}^{\phantom\dagger}\equiv\half\int\frac{d^2p}{(2\pi)^2}\int\frac{d^2p'}{(2\pi)^2}V_{\alpha\beta;\gamma\delta}(\b{p},\b{p}')c_{\b{p}\alpha}^\dag c_{\b{p}\beta}^{\phantom\dagger}c_{\b{p}'\gamma}^\dag c_{\b{p}'\delta}^{\phantom\dagger},
\end{align}
where the factor of two has been absorbed in a reduced interaction $V_{\alpha\beta;\gamma\delta}(\b{p},\b{p}')$ that only depends on two momenta $\b{p}$ and $\b{p}'$. Finally, since in Fermi liquid theory we focus on momenta near the Fermi surface, we can neglect the dependence of $V_{\alpha\beta;\gamma\delta}(\b{p},\b{p}')$ on the magnitudes $|\b{p}|$ and $|\b{p}'|$. One thus sets the interaction equal to its value on the (circular) Fermi surface,
\begin{align}\label{VFL}
V\approx\half\int\frac{d^2p}{(2\pi)^2}\int\frac{d^2p'}{(2\pi)^2}V_{\alpha\beta;\gamma\delta}(\hat{\b{p}},\hat{\b{p}}')c_{\b{p}\alpha}^\dag c_{\b{p}\beta}^{\phantom\dagger}c_{\b{p}'\gamma}^\dag c_{\b{p}'\delta}^{\phantom\dagger},
\end{align}
where $V_{\alpha\beta;\gamma\delta}(\hat{\b{p}},\hat{\b{p}}')\equiv V_{\alpha\beta;\gamma\delta}(p_F\hat{\b{p}},p_F\hat{\b{p}}')$ with $\hat{\b{p}},\hat{\b{p}}'$ unit vectors in the direction of $\b{p},\b{p}'$, and $p_F$ is the Fermi momentum. Since the quasiparticle matrix distribution function $n_\b{p}^{\alpha\beta}$ has the same symmetry properties as the expectation value $\langle c_{\b{p}\alpha}^\dag c_{\b{p}\beta}\rangle$, the form of the Landau functional (1) in the main text follows.

\section{Constraining the interaction by symmetries}
\label{sec:constrainsymm}

This section explains how to work out the most generic form of $V_{\alpha\beta;\gamma\delta}(\hat{\b{p}},\hat{\b{p}}')$ consistent with the symmetries of the problem.

Even before considering symmetries, particle indistinguishability (i.e., fermionic or bosonic statistics) gives us
\begin{align}
V_{\gamma\delta;\alpha\beta}(\hat{\b{p}}',\hat{\b{p}})=V_{\alpha\beta;\gamma\delta}(\hat{\b{p}},\hat{\b{p}}'),
\end{align}
and Hermiticity of $V$ gives us
\begin{align}
V^*_{\beta\alpha;\delta\gamma}(\hat{\b{p}},\hat{\b{p}}')=V_{\alpha\beta;\gamma\delta}(\hat{\b{p}},\hat{\b{p}}').
\end{align}
For each $\hat{\b{p}},\hat{\b{p}}'$, $V_{\alpha\beta;\gamma\delta}(\hat{\b{p}},\hat{\b{p}}')$ forms a $4\times 4$ matrix in spin indices. A natural basis on which we can expand this matrix is given by the tensor product of two sets of Pauli matrices $\sigma^\mu=(1,\bsigma)$, $\mu=0,1,2,3$ where $1$ denotes the $2\times 2$ identity matrix. We can write
\begin{align}\label{V1324munu}
V_{\alpha\beta;\gamma\delta}(\hat{\b{p}},\hat{\b{p}}')=\sum_{\mu\nu}V_{\mu\nu}(\hat{\b{p}},\hat{\b{p}}')\sigma^\mu_{\alpha\beta}\sigma^\nu_{\gamma\delta},
\end{align}
such that in virtue of Eq.~(\ref{VFL}), $V_{00}$ corresponds to charge-charge interactions, $V_{IJ}$ to spin-spin interactions, and $V_{0I},V_{I0}$ to spin-charge interactions. We will use uppercase indices $I,J$ for all three components $x,y,z$ of spin and lowercase indices $i,j$ for the in-plane components $x,y$. In this new basis, particle indistinguishability requires $V_{\mu\nu}(\hat{\b{p}},\hat{\b{p}}')=V_{\nu\mu}(\hat{\b{p}}',\hat{\b{p}})$ and Hermiticity requires $V_{\mu\nu}(\hat{\b{p}},\hat{\b{p}}')=V_{\mu\nu}^*(\hat{\b{p}},\hat{\b{p}}')$, i.e., the matrix $V_{\mu\nu}$ is real. Since the charge density and spin density are even and odd under time reversal, respectively, time-reversal symmetry requires
\begin{align}
V_{00}(-\hat{\b{p}},-\hat{\b{p}}')=V_{00}(\hat{\b{p}},\hat{\b{p}}'),\hspace{5mm}V_{IJ}(-\hat{\b{p}},-\hat{\b{p}}')=V_{IJ}(\hat{\b{p}},\hat{\b{p}}'),
\end{align}
for charge-charge and spin-spin interactions, and
\begin{align}
V_{0I}(-\hat{\b{p}},-\hat{\b{p}}')=-V_{0I}(\hat{\b{p}},\hat{\b{p}}'),\hspace{5mm}V_{I0}(-\hat{\b{p}},-\hat{\b{p}}')=-V_{I0}(\hat{\b{p}},\hat{\b{p}}'),
\end{align}
for spin-charge interactions. Note that $V_{0I}(\hat{\b{p}},\hat{\b{p}}')=V_{I0}(\hat{\b{p}}',\hat{\b{p}})$ from particle indistinguishability.

The interaction $V_{\mu\nu}(\hat{\b{p}},\hat{\b{p}}')$ can be expanded in angular momentum components,
\begin{align}\label{Vmunull}
V_{\mu\nu}(\hat{\b{p}},\hat{\b{p}}')=\sum_{l,l'=-\infty}^\infty V_{\mu\nu}^{ll'}e^{i(l\theta_\b{p}+l'\theta_{\b{p}'})},
\end{align}
where $\hat{\b{p}}=(\cos\theta_\b{p},\sin\theta_\b{p})$. Particle indistinguishability implies $V_{\mu\nu}^{ll'}=V_{\nu\mu}^{l'l}$ and Hermiticity implies $V_{\mu\nu}^{-l,-l'}=(V_{\mu\nu}^{ll'})^*$. For charge-charge and spin-spin interactions, time-reversal symmetry implies
\begin{align}
V_{00}^{ll'}=(-1)^{l+l'}V_{00}^{ll'},\hspace{5mm}V_{IJ}^{ll'}=(-1)^{l+l'}V_{IJ}^{ll'}.
\end{align}
In other words, $l+l'$ must be even for these coefficients to be nonzero, which is the same as saying that $l$ and $l'$ must have the same parity. Therefore $l'=l+2m$, $m\in\mathbb{Z}$, and we can write
\begin{align}\label{V00VijTRS}
V_{00}^{ll'}=V_{00}^{l,l+2m},\hspace{5mm}V_{IJ}^{ll'}=V_{IJ}^{l,l+2m},\hspace{5mm}m\in\mathbb{Z}.
\end{align}
For spin-charge interactions, time-reversal symmetry implies
\begin{align}
V_{0I}^{ll'}=-(-1)^{l+l'}V_{0I}^{ll'},\hspace{5mm}V_{I0}^{ll'}=-(-1)^{l+l'}V_{I0}^{ll'},
\end{align}
thus for these coefficients to be nonzero $l+l'$ must be odd, which is equivalent to saying that $l$ and $l'$ must have opposite parity. Therefore $l'=l+2m+1$, $m\in\mathbb{Z}$, and we have
\begin{align}\label{V0iTRS}
V_{0I}^{ll'}=V_{0I}^{l,l+2m+1},\hspace{5mm}V_{I0}^{ll'}=V_{I0}^{l,l+2m+1},\hspace{5mm}m\in\mathbb{Z}.
\end{align}

We now turn to implementing $SO(2)$ rotation symmetry, which is more subtle. From Eq.~(\ref{VFL}) and (\ref{V1324munu}) we can write
\begin{align}
V=\half\int\frac{d^2p}{(2\pi)^2}\int\frac{d^2p'}{(2\pi)^2}V_{\mu\nu}(\hat{\b{p}},\hat{\b{p}}')n_\mu(\b{p})n_\nu(\b{p}'),
\end{align}
where $n_0(\b{p})=n(\b{p})$ is the charge density and $n_I(\b{p})=2s_I(\b{p})$, $I=x,y,z$ is (twice) the spin density. Using $V_{0I}(\hat{\b{p}},\hat{\b{p}}')=V_{I0}(\hat{\b{p}}',\hat{\b{p}})$ from particle indistinguishability, we have
\begin{align}
V=V_{cc}+V_{sc}+V_{ss},
\end{align}
where
\begin{align}
V_{cc}&=\half\int\frac{d^2p}{(2\pi)^2}\int\frac{d^2p'}{(2\pi)^2}V_{00}(\hat{\b{p}},\hat{\b{p}}')n(\b{p})n(\b{p}'),\label{VccDEF}\\
V_{sc}&=2\int\frac{d^2p}{(2\pi)^2}\int\frac{d^2p'}{(2\pi)^2}V_{0I}(\hat{\b{p}},\hat{\b{p}}')n(\b{p})s_I(\b{p}'),\label{VscDEF}\\
V_{ss}&=2\int\frac{d^2p}{(2\pi)^2}\int\frac{d^2p'}{(2\pi)^2}V_{IJ}(\hat{\b{p}},\hat{\b{p}}')s_I(\b{p})s_J(\b{p}'),\label{VssDEF}
\end{align}
are the charge-charge, spin-charge, and spin-spin interactions, respectively. Using Eq.~(\ref{Vmunull}) and the constraints from time-reversal symmetry (\ref{V00VijTRS}) and (\ref{V0iTRS}), we have
\begin{align}
V_{cc}&=\half\sum_{lm}\int\frac{d^2p}{(2\pi)^2}\int\frac{d^2p'}{(2\pi)^2}e^{il\theta_\b{p}+i(l+2m)\theta_{\b{p}'}}V_{00}^{l,l+2m}
n(\b{p})n(\b{p}'),\label{VccbeforeSO2}\\
V_{sc}&=2\sum_{lm}\int\frac{d^2p}{(2\pi)^2}\int\frac{d^2p'}{(2\pi)^2}e^{il\theta_\b{p}+i(l+2m+1)\theta_{\b{p}'}}V_{0I}^{l,l+2m+1}
n(\b{p})s_I(\b{p}'),\label{VscbeforeSO2}\\
V_{ss}&=2\sum_{lm}\int\frac{d^2p}{(2\pi)^2}\int\frac{d^2p'}{(2\pi)^2}e^{il\theta_\b{p}+i(l+2m)\theta_{\b{p}'}}V_{IJ}^{l,l+2m}
s_I(\b{p})s_J(\b{p}').\label{VssbeforeSO2}
\end{align}
To implement $SO(2)$ symmetry, we note that the annihilation operator transforms as
\begin{align}
R(\varphi)c_{\b{p}\alpha}R(\varphi)^{-1}=\left(e^{-i\varphi\sigma_3/2}\right)_{\alpha\alpha'}c_{R_\varphi^{-1}\b{p},\alpha'},
\end{align}
where $R(\varphi)$ on the left-hand side is the rotation operator, and $R_\varphi$ on the right-hand side is the usual $2\times 2$ rotation matrix
\begin{align}\label{RotMatrix}
R_\varphi=\left(\begin{array}{cc}
\cos\varphi & -\sin\varphi \\
\sin\varphi & \cos\varphi
\end{array}\right).
\end{align}
This implies that the charge and spin densities transform as
\begin{align}\label{SO2chargespin}
R(\varphi)n(\b{p})R(\varphi)^{-1}&=n(R_\varphi^{-1}\b{p}),\nonumber\\
R(\varphi)s_i(\b{p})R(\varphi)^{-1}&=R_\varphi^{ij}s_j(R_\varphi^{-1}\b{p}),\text{ for }i,j=x,y,\nonumber\\
R(\varphi)s_z(\b{p})R(\varphi)^{-1}&=s_z(R_\varphi^{-1}\b{p}).
\end{align}
We require that $R(\varphi)VR(\varphi)^{-1}=V$. From Eq.~(\ref{SO2chargespin}) it is clear that the charge-charge, spin-charge, and spin-spin interaction terms in Eq.~(\ref{VccbeforeSO2}), (\ref{VscbeforeSO2}), and (\ref{VssbeforeSO2}) will transform into themselves under $SO(2)$ rotations, and we can look at each term in turn.

\subsection{Charge-charge interaction}

For the charge-charge interaction, requiring $R(\varphi)V_{cc}R(\varphi)^{-1}=V_{cc}$ implies
\begin{align}
\sum_{lm} e^{i2(l+m)\varphi}&\int\frac{d^2p}{(2\pi)^2}\int\frac{d^2p'}{(2\pi)^2}e^{il\theta_\b{p}+i(l+2m)\theta_{\b{p}'}}V_{00}^{l,l+2m}n(\b{p})n(\b{p}')\nonumber\\
&=\sum_{lm}\int\frac{d^2p}{(2\pi)^2}\int\frac{d^2p'}{(2\pi)^2}e^{il\theta_\b{p}+i(l+2m)\theta_{\b{p}'}}V_{00}^{l,l+2m}
n(\b{p})n(\b{p}'),
\end{align}
for arbitrary $\varphi$, which implies the constraint $l=-m$. We therefore obtain
\begin{align}
V_{cc}=\half\sum_m\int\frac{d^2p}{(2\pi)^2}\int\frac{d^2p'}{(2\pi)^2}e^{im\theta_{\b{p}\b{p}'}}V_{00}^{-m,m}n(\b{p})n(\b{p}'),
\end{align}
where $\theta_{\b{p}\b{p}'}\equiv\theta_{\b{p}'}-\theta_\b{p}$ is the angle between $\b{p}$ and $\b{p}'$. The charge-charge interaction matrix element $V_{00}(\hat{\b{p}},\hat{\b{p}}')$ can only depend on this relative angle, i.e.,
\begin{align}
V_{00}(\hat{\b{p}},\hat{\b{p}}')=V_{00}(\theta_{\b{p}\b{p}'})=\sum_m V_{00}^{-m,m'}e^{im\theta_{\b{p}\b{p}'}}.
\end{align}
In fact, because of particle indistinguishability this matrix element is symmetric under $\hat{\b{p}}\leftrightarrow\hat{\b{p}}'$ and therefore depends only on the cosine of $\theta_{\b{p}\b{p}'}$,
\begin{align}
V_{00}(\theta_{\b{p}\b{p}'})=V_{00}^{00}+2\sum_{m=1}^\infty V_{00}^{-m,m}\cos m\theta_{\b{p}\b{p}'}.
\end{align}
Therefore the charge-charge interaction is
\begin{align}\label{Vccreal}
V_{cc}=\half\int\frac{d^2p}{(2\pi)^2}\int\frac{d^2p'}{(2\pi)^2}\left(V_{00}^{00}+2\sum_{m=1}^\infty V_{00}^{-m,m}\cos m\theta_{\b{p}\b{p}'}\right)n(\b{p})n(\b{p}'),
\end{align}
where the $V_{00}^{-m,m}$, $m=0,1,2,\ldots$ are real coefficients.

\subsection{Spin-charge interaction}

For the spin-charge interaction, it is convenient to first separate the $x,y$ components of spin from the $z$ component, as they have different transformation properties under rotations,
\begin{align}
V_{sc}&=2\sum_{lm}\int\frac{d^2p}{(2\pi)^2}\int\frac{d^2p'}{(2\pi)^2}e^{il\theta_\b{p}+i(l+2m+1)\theta_{\b{p}'}}\left(V_{0i}^{l,l+2m+1}n(\b{p})s_i(\b{p}')
+V_{0z}^{l,l+2m+1}n(\b{p})s_z(\b{p}')\right).
\end{align}
Requiring $R(\varphi)V_{sc}R(\varphi)^{-1}=V_{sc}$ implies
\begin{align}
\sum_{lm} e^{i[2(l+m)+1]\varphi}&\int\frac{d^2p}{(2\pi)^2}\int\frac{d^2p'}{(2\pi)^2}e^{il\theta_\b{p}+i(l+2m+1)\theta_{\b{p}'}}\left(V_{0i}^{l,l+2m+1}R_\varphi^{ij}n(\b{p})s_j(\b{p}')
+V_{0z}^{l,l+2m+1}n(\b{p})s_z(\b{p}')\right)\nonumber\\
&=\sum_{lm}\int\frac{d^2p}{(2\pi)^2}\int\frac{d^2p'}{(2\pi)^2}e^{il\theta_\b{p}+i(l+2m+1)\theta_{\b{p}'}}\left(V_{0i}^{l,l+2m+1}n(\b{p})s_i(\b{p}')
+V_{0z}^{l,l+2m+1}n(\b{p})s_z(\b{p}')\right).
\end{align}
For the $0z$ component of the interaction, this implies $V_{0z}^{l,l+2m+1}=e^{i[2(l+m)+1]\varphi}V_{0z}^{l,l+2m+1}$ for arbitrary $\varphi$, which would require $2(l+m)+1=0$. This is impossible since $l,m$ are integers, hence $V_{0z}^{l,l+2m+1}=0$. For the $0x$ and $0y$ components, we have
\begin{align}
e^{i[2(l+m)+1]\varphi}V_{0i}^{l,l+2m+1}R_\varphi^{ij}=V_{0j}^{l,l+2m+1},
\end{align}
which is equivalent to
\begin{align}\label{EigenEqV0i}
R_\varphi^{ij}V_{0j}^{l,l+2m+1}=e^{i[2(l+m)+1]\varphi}V_{0i}^{l,l+2m+1}.
\end{align}
In other words, $V_{0i}^{l,l+2m+1}$ must be an eigenvector of the rotation matrix $R_\varphi$ with eigenvalue $e^{i[2(l+m)+1]\varphi}$. The eigenvalues and eigenvectors of the rotation matrix (\ref{RotMatrix}) are given by
\begin{align}\label{EigRotMatrix}
e^{\pm i\varphi},\hspace{5mm}\frac{1}{\sqrt{2}}\left(\begin{array}{c}
\pm i\\
1
\end{array}\right),
\end{align}
respectively. For the eigenvalue $e^{i\varphi}$, this implies that $2(l+m)+1=1$ which is satisfied by $l=-m$. The form of the corresponding eigenvector implies that
\begin{align}
V_{0y}^{-m,m+1}=-iV_{0x}^{-m,m+1}.
\end{align}
For the eigenvalue $e^{-i\varphi}$, this implies that $2(l+m)+1=-1$ which is satisfied by $l=-m-1$. The form of the corresponding eigenvector implies that
\begin{align}
V_{0y}^{-m-1,m}=iV_{0x}^{-m-1,m}.
\end{align}
Considering these two possible values of $l$, the spin-charge interaction becomes
\begin{align}\label{Vscnotreal}
V_{sc}&=2\sum_m\int\frac{d^2p}{(2\pi)^2}\int\frac{d^2p'}{(2\pi)^2}e^{im\theta_{\b{p}\b{p}'}}\left[V_{0x}^{-m,m+1}n(\b{p})e^{i\theta_{\b{p}'}}s_-(\b{p}')+V_{0x}^{-m,m-1}n(\b{p})e^{-i\theta_{\b{p}'}}s_+(\b{p}')\right]\nonumber\\
&=2\sum_m\int\frac{d^2p}{(2\pi)^2}\int\frac{d^2p'}{(2\pi)^2}e^{im\theta_{\b{p}\b{p}'}}V_{0x}^{-m,m+1}n(\b{p})e^{i\theta_{\b{p}'}}s_-(\b{p}')+\mathrm{H.c.}
\end{align}
Unlike in a standard Fermi liquid, here there exists a time-reversal and rotationally invariant spin-charge interaction.

\subsection{Spin-spin interaction}

Defining a 3D rotation matrix
\begin{align}
\mathcal{R}_\varphi\equiv\left(\begin{array}{cc}
R_\varphi & 0 \\
0 & 1
\end{array}\right),
\end{align}
and denoting the $3\times 3$ spin-spin interaction matrix $V^{l,l+2m}_{IJ}$ by $\b{V}^{l,l+2m}$ and the 3-component spin vector $s_I(\b{p})$ by $\b{s}(\b{p})$, we can write the spin-spin interaction term (\ref{VssbeforeSO2}) as
\begin{align}
V_{ss}&=2\sum_{lm}\int\frac{d^2p}{(2\pi)^2}\int\frac{d^2p'}{(2\pi)^2}e^{il\theta_\b{p}+i(l+2m)\theta_{\b{p}'}}\b{s}(\b{p})^T\b{V}^{l,l+2m}\b{s}(\b{p}').
\end{align}
By virtue of Eq.~(\ref{SO2chargespin}), $\b{s}(\b{p})$ transforms under rotations as
\begin{align}
R(\varphi)\b{s}(\b{p})R(\varphi)^{-1}=\mathcal{R}_\varphi\b{s}(R_\varphi^{-1}\b{p}).
\end{align}
Therefore, requiring $R(\varphi)V_{ss}R(\varphi)^{-1}=V_{ss}$ implies
\begin{align}
\sum_{lm}e^{i2(l+m)\varphi}&\int\frac{d^2p}{(2\pi)^2}\int\frac{d^2p'}{(2\pi)^2}e^{il\theta_\b{p}+i(l+2m)\theta_{\b{p}'}}\b{s}(\b{p})^T\mathcal{R}_\varphi^{-1}\b{V}^{l,l+2m}\mathcal{R}_\varphi\b{s}(\b{p}')\nonumber\\
&=\sum_{lm}\int\frac{d^2p}{(2\pi)^2}\int\frac{d^2p'}{(2\pi)^2}e^{il\theta_\b{p}+i(l+2m)\theta_{\b{p}'}}\b{s}(\b{p})^T\b{V}^{l,l+2m}\b{s}(\b{p}').
\end{align}
This implies $e^{i2(l+m)\varphi}\mathcal{R}_\varphi^{-1}\b{V}^{l,l+2m}\mathcal{R}_\varphi=\b{V}^{l,l+2m}$, or equivalently
\begin{align}\label{RotCondVss}
\mathcal{R}_\varphi\b{V}^{l,l+2m}\mathcal{R}_\varphi^{-1}
=e^{i2(l+m)\varphi}\b{V}^{l,l+2m},
\end{align}
for arbitrary $\varphi$. Separating the $3\times 3$ matrix into in-plane ($x,y$) and $z$ components,
\begin{align}
\b{V}^{l,l+2m}=\left(\begin{array}{cc}
\b{V}^{l,l+2m}_\parallel & \b{V}_{\parallel,z}^{l,l+2m} \\
\b{V}^{l,l+2m}_{z,\parallel} & V^{l,l+2m}_{zz}
\end{array}
\right),
\end{align}
where $\b{V}^{l,l+2m}_\parallel$ is a $2\times 2$ matrix, $\b{V}_{\parallel,z}^{l,l+2m}$ is a $2\times 1$ column vector, and $\b{V}^{l,l+2m}_{z,\parallel}$ is a $1\times 2$ row vector, condition (\ref{RotCondVss}) translates into the four conditions
\begin{align}
R_\varphi\b{V}^{l,l+2m}_\parallel R_\varphi^{-1}&=e^{i2(l+m)\varphi}\b{V}^{l,l+2m}_\parallel,\label{RotCondVinplane}\\
R_\varphi\b{V}_{\parallel,z}^{l,l+2m}&=e^{i2(l+m)\varphi}\b{V}_{\parallel,z}^{l,l+2m},\label{RotCondVparaz}\\
\b{V}^{l,l+2m}_{z,\parallel}R_\varphi^{-1}&=e^{i2(l+m)\varphi}\b{V}^{l,l+2m}_{z,\parallel},\label{RotCondVzpara}\\
V^{l,l+2m}_{zz}&=e^{i2(l+m)\varphi}V^{l,l+2m}_{zz}.\label{RotCondVzz}
\end{align}
Beginning with the simplest condition, Eq.~(\ref{RotCondVzz}) requires that $l=-m$, hence the $zz$ part of the spin-spin interaction becomes
\begin{align}
2\sum_{m}\int\frac{d^2p}{(2\pi)^2}\int\frac{d^2p'}{(2\pi)^2}e^{im\theta_{\b{p}\b{p}'}}V_{zz}^{-m,m}s_z(\b{p})s_z(\b{p}').
\end{align}
Similar to Eq.~(\ref{EigenEqV0i}), Eq.~(\ref{RotCondVparaz}) requires that $\b{V}_{\parallel,z}^{l,l+2m}$ be an eigenvector of the 2D rotation matrix $R_\varphi$ with eigenvalue $e^{i2(l+m)\varphi}$. As seen before, the eigenvalues of $R_\varphi$ are $e^{\pm i\varphi}$. Since $l,m$ are integers, condition (\ref{RotCondVparaz}) can only be satisfied if $\b{V}_{\parallel,z}^{l,l+2m}=0$, thus $V^{l,l+2m}_{xz}=V^{l,l+2m}_{yz}=0$. Likewise, Eq.~(\ref{RotCondVzpara}) is equivalent to the eigenvalue condition
\begin{align}
R_\varphi(\b{V}^{l,l+2m}_{z,\parallel})^T=e^{i2(l+m)\varphi}(\b{V}^{l,l+2m}_{z,\parallel})^T,
\end{align}
which can only be satisfied if $(\b{V}^{l,l+2m}_{z,\parallel})^T=0$, thus $V^{l,l+2m}_{zx}=V^{l,l+2m}_{zy}=0$.

Finally, Eq.~(\ref{RotCondVinplane}) can also be converted to an eigenvalue condition by expanding the $2\times 2$ matrix $\b{V}^{l,l+2m}_\parallel$ on the basis of Pauli matrices plus the identity matrix,
\begin{align}
\b{V}^{l,l+2m}_\parallel=\tilde{V}_\alpha^{l,l+2m}\sigma^\alpha,
\end{align}
with $\alpha=0,1,2,3$. The rotation matrix $R_\varphi$ can be written as $R_\varphi=\sigma^0\cos\varphi-i\sigma^2\sin\varphi$, which gives the following transformation properties for the Pauli matrices,
\begin{align}
R_\varphi\sigma^\alpha R_\varphi^{-1}=\sigma^\alpha\left(\cos^2\varphi+(-1)^\alpha\sin^2\varphi\right)-2(1-\delta_{\alpha 0})\epsilon^{\alpha 2\beta}\sigma^\beta\sin\varphi\cos\varphi.
\end{align}
Using this property, Eq.~(\ref{RotCondVinplane}) is equivalent to the following condition,
\begin{align}
&\tilde{V}_0^{l,l+2m}\sigma^0+\left(\tilde{V}^{l,l+2m}_1\cos 2\varphi+\tilde{V}^{l,l+2m}_3\sin 2\varphi\right)\sigma^1+\tilde{V}^{l,l+2m}_2\sigma^2+\left(\tilde{V}^{l,l+2m}_3\cos 2\varphi-\tilde{V}^{l,l+2m}_1\sin 2\varphi\right)\sigma^3\nonumber\\
&=e^{i2(l+m)\varphi}\left(\tilde{V}_0^{l,l+2m}\sigma^0+
\tilde{V}_1^{l,l+2m}\sigma^1+\tilde{V}_2^{l,l+2m}\sigma^2
+\tilde{V}_3^{l,l+2m}\sigma^3\right).
\end{align}
The $\sigma^0$ and $\sigma^2$ terms give the conditions
\begin{align}\label{RotCond02}
\tilde{V}_0^{l,l+2m}=\tilde{V}_0^{-m,m}\delta_{l+m,0},\hspace{5mm}\tilde{V}_2^{l,l+2m}=\tilde{V}_2^{-m,m}\delta_{l+m,0},
\end{align}
while the $\sigma^1$ and $\sigma^3$ terms give the condition
\begin{align}
R_{2\varphi}\left(\begin{array}{c}
\tilde{V}_3^{l,l+2m} \\
\tilde{V}_1^{l,l+2m}\end{array}\right)
=e^{i2(l+m)\varphi}\left(\begin{array}{c}
\tilde{V}_3^{l,l+2m} \\
\tilde{V}_1^{l,l+2m}\end{array}\right).
\end{align}
This is again an eigenvalue condition, but this time for the rotation matrix with angle $2\varphi$ which has eigenvalues $e^{\pm i2\varphi}$. This implies that $l+m=\pm 1$, or $l=-m\pm 1$. The eigenvectors, however, are the same as in Eq.~(\ref{EigRotMatrix}). For $l=-m+1$, we have
\begin{align}\label{RotCond13p}
\tilde{V}_3^{-m+1,m+1}=i\tilde{V}_1^{-m+1,m+1},
\end{align}
while for $l=-m-1$, we have
\begin{align}\label{RotCond13m}
\tilde{V}_3^{-m-1,m-1}=-i\tilde{V}_1^{-m-1,m-1}.
\end{align}
The complete spin-spin interaction is therefore given by
\begin{align}
V_{ss}=2\sum_m\int\frac{d^2p}{(2\pi)^2}\int\frac{d^2p'}{(2\pi)^2}\left(\sum_l e^{il\theta_\b{p}+i(l+2m)\theta_{\b{p}'}}\tilde{V}^{l,l+2m}_\alpha\sigma_{ij}^\alpha s_i(\b{p})s_j(\b{p}')+e^{im\theta_{\b{p}\b{p}'}}V_{zz}^{-m,m}s_z(\b{p})s_z(\b{p}')\right),
\end{align}
which, upon substituting Eq.~(\ref{RotCond02}), (\ref{RotCond13p}), and (\ref{RotCond13m}), yields
\begin{align}\label{Vssnotreal}
V_{ss}=2\sum_m\int\frac{d^2p}{(2\pi)^2}\int\frac{d^2p'}{(2\pi)^2}e^{im\theta_{\b{p}\b{p}'}}\Bigl[&\tilde{V}^{-m,m}_0(s_xs_x'+s_ys_y')+V_{zz}^{-m,m}s_zs_z'
-i\tilde{V}_2^{-m,m}(s_xs_y'-s_ys_x')\nonumber\\
&+i\tilde{V}_1^{-m+1,m+1}\left(e^{i\theta_\b{p}}s_-\right)\left(e^{i\theta_{\b{p}'}}s_-'\right)
-i\tilde{V}_1^{-m-1,m-1}\left(e^{-i\theta_\b{p}}s_+\right)\left(e^{-i\theta_{\b{p}'}}s_+'\right)\Bigr],
\end{align}
where we denote $s_I\equiv s_I(\b{p})$ and $s_I'\equiv s_I(\b{p}')$. The first two terms correspond to an XXZ interaction, the third term to a Dzyaloshinskii-Moriya interaction which reflects the presence of spin-orbit coupling in the system, and the last two terms to anisotropic spin-spin interactions similar to those found in compass models,\cite{nussinov2015} but with a continuous rather than discrete symmetry. These terms are not invariant under separate spatial and spin rotations, but only under a simultaneous rotation in spin space and real space.

\section{Landau parameters}
\label{sec:landau}

Landau parameters are the real coefficients of Hermitian interaction terms. The charge-charge interaction (\ref{Vccreal}) is already in this form, but we wish to write the spin-charge interaction (\ref{Vscnotreal}) and the spin-spin interaction (\ref{Vssnotreal}) in this form as well. We first consider the spin-charge interaction. Splitting $V_{0x}^{-m,m+1}$ into real and imaginary parts,
\begin{align}
V_{0x}^{-m,m+1}=\tilde{V}_m'+i\tilde{V}_m'',
\end{align}
$V_{sc}$ can be written as
\begin{align}
V_{sc}=4\int\frac{d^2p}{(2\pi)^2}\int\frac{d^2p'}{(2\pi)^2}\Biggl\{&\left[\tilde{V}_0'+\sum_{m=1}^\infty\left((\tilde{V}'_m+\tilde{V}_{-m}')\cos m\theta_{\b{p}\b{p}'}
-(\tilde{V}''_m-\tilde{V}_{-m}'')\sin m\theta_{\b{p}\b{p}'}\right)\right]n(\b{p})\hat{\b{p}}'\cdot\b{s}'\nonumber\\
&+\left[\tilde{V}_0''+\sum_{m=1}^\infty\left((\tilde{V}''_m+\tilde{V}_{-m}'')\cos m\theta_{\b{p}\b{p}'}
+(\tilde{V}'_m-\tilde{V}_{-m}')\sin m\theta_{\b{p}\b{p}'}\right)\right]n(\b{p})\hat{\b{p}}'\times\b{s}'\Biggr\},
\end{align}
where we have used the fact that $e^{\pm i\theta_\b{p}}=\hat{p}_x\pm i\hat{p}_y$, and we denote $\hat{\b{z}}\cdot(\b{a}\times\b{b})\equiv\b{a}\times\b{b}$ for simplicity. For the spin-spin interaction, particle indistinguishability implies that $\tilde{V}_0^{-m,m}$, $V_{zz}^{-m,m}$, and $\tilde{V}_2^{-m,m}$ are real, while $(\tilde{V}_1^{-m+1,m+1})^*=\tilde{V}_1^{-m-1,m-1}$. Splitting $\tilde{V}_1^{-m+1,m+1}$ into real and imaginary parts,
\begin{align}
\tilde{V}_1^{-m+1,m+1}=(\tilde{V}_1^m)'+i(\tilde{V}_1^m)'',
\end{align}
the spin-spin interaction can be written as
\begin{align}
V_{ss}=2\int\frac{d^2p}{(2\pi)^2}\int\frac{d^2p'}{(2\pi)^2}\Biggl[&\left(\tilde{V}_0^{00}+2\sum_{m=1}^\infty\tilde{V}_0^{-m,m}\cos m\theta_{\b{p}\b{p}'}\right)\b{s}\cdot\b{s}'+\left(V_{zz}^{00}+2\sum_{m=1}^\infty V_{zz}^{-m,m}\cos m\theta_{\b{p}\b{p}'}\right)s_zs_z'\nonumber\\
&+2\sum_{m=1}^\infty\tilde{V}_2^{-m,m}\sin m\theta_{\b{p}\b{p}'}\b{s}\times\b{s}'\nonumber\\
&+2\left((\tilde{V}_1^0)'+2\sum_{m=1}^\infty(\tilde{V}_1^m)'\cos m\theta_{\b{p}\b{p}'}\right)\left((\hat{\b{p}}\cdot\b{s})(\hat{\b{p}}'\times\b{s}')+(\hat{\b{p}}\times\b{s})(\hat{\b{p}}'\cdot\b{s}')\right)\nonumber\\
&-2\left((\tilde{V}_1^0)''+2\sum_{m=1}^\infty(\tilde{V}_1^m)''\cos m\theta_{\b{p}\b{p}'}\right)\left((\hat{\b{p}}\cdot\b{s})(\hat{\b{p}}'\cdot\b{s}')-(\hat{\b{p}}\times\b{s})(\hat{\b{p}}'\times\b{s}')\right)\Biggr].
\end{align}
Considering the full interaction term $V=V_{cc}+V_{sc}+V_{ss}$, for each $m$ there are ten independent real coefficients, hence ten Landau parameters. We define one charge-charge Landau parameter $f_m^{cc}$,
\begin{align}
f^{cc}_m=\left\{\begin{array}{cc}
V_{00}^{00}, & m=0 \\
2V_{00}^{-m,m}, & m=1,2,3,\ldots
\end{array}\right.
\end{align}
four spin-charge Landau parameters $f_m^{sc,1},\ldots,f_m^{sc,4}$,
\begin{align}
f^{sc,1}_m&=\left\{\begin{array}{cc}
4\tilde{V}_0', & m=0 \\
4(\tilde{V}_m'+\tilde{V}_{-m}'), & m=1,2,3,\ldots
\end{array}\right.\\
f^{sc,2}_m&=\left\{\begin{array}{cc}
0, & m=0 \\
-4(\tilde{V}_m''-\tilde{V}_{-m}''), & m=1,2,3,\ldots
\end{array}\right.\\
f^{sc,3}_m&=\left\{\begin{array}{cc}
4\tilde{V}_0'', & m=0 \\
4(\tilde{V}_m''+\tilde{V}_{-m}''), & m=1,2,3,\ldots
\end{array}\right.\\
f^{sc,4}_m&=\left\{\begin{array}{cc}
0, & m=0 \\
4(\tilde{V}_m'-\tilde{V}_{-m}'), & m=1,2,3,\ldots
\end{array}\right.
\end{align}
and five spin-spin Landau parameters $f_m^{ss,1},\ldots,f_m^{ss,5}$,
\begin{align}
f^{ss,1}_m&=\left\{\begin{array}{cc}
4\tilde{V}_0^{00}, & m=0 \\
8\tilde{V}_0^{-m,m}, & m=1,2,3,\ldots
\end{array}\right.\\
f^{ss,2}_m&=\left\{\begin{array}{cc}
4V_{zz}^{00}, & m=0 \\
8V_{zz}^{-m,m}, & m=1,2,3,\ldots
\end{array}\right.\\
f^{ss,3}_m&=\left\{\begin{array}{cc}
0, & m=0 \\
8\tilde{V}_2^{-m,m}, & m=1,2,3,\ldots
\end{array}\right.\\
f^{ss,4}_m&=\left\{\begin{array}{cc}
8(\tilde{V}_1^0)', & m=0 \\
16(\tilde{V}_1^m)', & m=1,2,3,\ldots
\end{array}\right.\\
f^{ss,5}_m&=\left\{\begin{array}{cc}
-8(\tilde{V}_1^0)'', & m=0 \\
-16(\tilde{V}_1^m)'', & m=1,2,3,\ldots
\end{array}\right.
\end{align}
In terms of these Landau parameters, the interaction terms can be written as
\begin{align}\label{VccFINAL}
V_{cc}&=\half\sum_{m=0}^\infty\int\frac{d^2p}{(2\pi)^2}\int\frac{d^2p'}{(2\pi)^2}f^{cc}_m\cos m\theta_{\b{p}\b{p}'}n(\b{p})n(\b{p}'),
\end{align}
for the charge-charge interaction,
\begin{align}\label{VscFINAL}
V_{sc}=\sum_{m=0}^\infty\int\frac{d^2p}{(2\pi)^2}\int\frac{d^2p'}{(2\pi)^2}\Bigl[&\left(f^{sc,1}_m\cos m\theta_{\b{p}\b{p}'}+f^{sc,2}_m\sin m\theta_{\b{p}\b{p}'}\right)n(\b{p})\hat{\b{p}}'\cdot\b{s}(\b{p}')\nonumber\\
&+\left(f^{sc,3}_m\cos m\theta_{\b{p}\b{p}'}+f^{sc,4}_m\sin m\theta_{\b{p}\b{p}'}\right)n(\b{p})\hat{\b{p}}'\times\b{s}(\b{p}')\Bigr],
\end{align}
for the spin-charge interaction, and
\begin{align}\label{VssFINAL}
V_{ss}=\half\sum_{m=0}^\infty\int\frac{d^2p}{(2\pi)^2}\int\frac{d^2p'}{(2\pi)^2}\Bigl\{&f^{ss,1}_m\cos m\theta_{\b{p}\b{p}'}\b{s}(\b{p})\cdot\b{s}(\b{p}')
+f^{ss,2}_m\cos m\theta_{\b{p}\b{p}'}s_z(\b{p})s_z(\b{p}')+f^{ss,3}_m\sin m\theta_{\b{p}\b{p}'}\b{s}(\b{p})\times\b{s}(\b{p}')\nonumber\\
&+f^{ss,4}_m\cos m\theta_{\b{p}\b{p}'}\left[\left(\hat{\b{p}}\cdot\b{s}(\b{p})\right)\left(\hat{\b{p}}'\times\b{s}(\b{p}')\right)+\left(\hat{\b{p}}\times\b{s}(\b{p})\right)\left(\hat{\b{p}}'\cdot\b{s}(\b{p}')\right)\right]\nonumber\\
&+f^{ss,5}_m\cos m\theta_{\b{p}\b{p}'}\left[\left(\hat{\b{p}}\cdot\b{s}(\b{p})\right)\left(\hat{\b{p}}'\cdot\b{s}(\b{p}')\right)-\left(\hat{\b{p}}\times\b{s}(\b{p})\right)\left(\hat{\b{p}}'\times\b{s}(\b{p}')\right)\right]\Bigr\},
\end{align}
for the spin-spin interaction. Interpreted as quasiparticle interaction terms in a Landau functional, Eq.~(\ref{VccFINAL})-(\ref{VssFINAL}) correspond to Eq.~(5)-(7) of the main text.

\subsection{Projected Landau parameters}

In this section we explain how to derive the projected Landau functional [Eq.~(8) of the main text] from the unprojected theory we have just described. The starting point is to drop the negative helicity part in the expressions for the fermion operators,
\begin{align}
c_{\b{p}\ua}&=\frac{ie^{-i\theta_\b{p}}}{\sqrt{2}}(\psi_{\b{p}+}+\psi_{\b{p}-})\approx\frac{ie^{-i\theta_\b{p}}}{\sqrt{2}}\psi_{\b{p}+},\\
c_{\b{p}\da}&=\frac{1}{\sqrt{2}}(\psi_{\b{p}+}-\psi_{\b{p}-})\approx\frac{1}{\sqrt{2}}\psi_{\b{p}+},
\end{align}
which can be expressed as $c_{\b{p}\sigma}\approx\eta_{\hat{\b{p}}\sigma}\psi_\b{p}$ where the $c$-number spinor $\eta_{\hat{\b{p}}}=\frac{1}{\sqrt{2}}(ie^{-i\theta_\b{p}},1)$ obeys $\eta_{\hat{\b{p}}}^\dag\eta_{\hat{\b{p}}}^{\phantom\dagger}=1$, and we define the effectively spinless fermion operator $\psi_\b{p}\equiv\psi_{\b{p}+}$. One then substitutes this expression for $c_{\b{p}\sigma}$ into the interaction Hamiltonian (\ref{VFL}). This produces a projected interaction Hamiltonian $\bar{V}$,
\begin{align}\label{RedSpinlessInt}
\bar{V}=\half\int\frac{d^2p}{(2\pi)^2}\int\frac{d^2p'}{(2\pi)^2}\bar{V}(\hat{\b{p}},\hat{\b{p}}')\psi_{\b{p}}^\dag\psi_{\b{p}}^{\phantom\dagger}\psi_{\b{p}'}^\dag\psi_{\b{p}'}^{\phantom\dagger},
\end{align}
where the projected matrix element $\bar{V}(\hat{\b{p}},\hat{\b{p}}')$ is given in terms of the unprojected ones by
\begin{align}
\bar{V}(\hat{\b{p}},\hat{\b{p}}')\equiv V_{\alpha\gamma;\beta\delta}(\hat{\b{p}},\hat{\b{p}}')
\eta_{\hat{\b{p}}\alpha}^*\eta_{\hat{\b{p}}\beta}^{\phantom *}
\eta_{\hat{\b{p}}'\gamma}^*\eta_{\hat{\b{p}}'\delta}^{\phantom *}
=\sum_{\mu\nu}V_{\mu\nu}(\hat{\b{p}},\hat{\b{p}}')(\eta_{\hat{\b{p}}}^\dag\sigma^\mu\eta_{\hat{\b{p}}}^{\phantom\dagger})(\eta_{\hat{\b{p}}'}^\dag\sigma^\nu\eta_{\hat{\b{p}}'}^{\phantom\dagger}),
\end{align}
where we used Eq.~(\ref{V1324munu}). The quantity $\eta_{\hat{\b{p}}}^\dag\sigma^\mu\eta_{\hat{\b{p}}}^{\phantom\dagger}$ can be thought of as the expectation value of $\sigma^\mu$ in the single-particle eigenstate at $\hat{\b{p}}$ on the Fermi surface. We have $\eta_{\hat{\b{p}}}^\dag\sigma^0\eta_{\hat{\b{p}}}^{\phantom\dagger}=1$ which corresponds to a particle number of one, $\eta_{\hat{\b{p}}}^\dag\sigma^3\eta_{\hat{\b{p}}}^{\phantom\dagger}=0$ which indicates that spin polarization on the Fermi surface is entirely in-plane, and $\eta_{\hat{\b{p}}}^\dag\sigma^i\eta_{\hat{\b{p}}}^{\phantom\dagger}=\epsilon_{ij}\hat{p}_j$, $i,j=1,2$, which indicates that spin is perpendicular to momentum everywhere on the Fermi surface, i.e., spin-momentum locking. Using the fact that $V_{0i}(\hat{\b{p}},\hat{\b{p}}')=V_{i0}(\hat{\b{p}}',\hat{\b{p}})$, and the fact that the interaction (\ref{RedSpinlessInt}) must be symmetric under $\hat{\b{p}}\leftrightarrow\hat{\b{p}}'$, we have
\begin{align}\label{VbarSum}
\bar{V}(\hat{\b{p}},\hat{\b{p}}')=V_{00}(\hat{\b{p}},\hat{\b{p}}')+2V_{0i}(\hat{\b{p}},\hat{\b{p}}')\epsilon_{ij}\hat{p}_j'+V_{ij}(\hat{\b{p}},\hat{\b{p}}')\epsilon_{ik}\epsilon_{jl}\hat{p}_k\hat{p}_l'.
\end{align}
The $V_{zz}$ component of the spin-spin interaction does not enter since spins on the Fermi surface are entirely in-plane. We can now read off $V_{00}(\hat{\b{p}},\hat{\b{p}}')$, $V_{0i}(\hat{\b{p}},\hat{\b{p}}')$, and $V_{ij}(\hat{\b{p}},\hat{\b{p}}')$ from the interaction terms (\ref{VccFINAL}), (\ref{VscFINAL}), and (\ref{VssFINAL}), and their original definitions (\ref{VccDEF}), (\ref{VscDEF}), and (\ref{VssDEF}), respectively. We find
\begin{align}
V_{00}(\hat{\b{p}},\hat{\b{p}}')&=\sum_{m=0}^\infty f^{cc}_m\cos m\theta_{\b{p}\b{p}'},\\
V_{0i}(\hat{\b{p}},\hat{\b{p}}')&=\frac{1}{2}\sum_{m=0}^\infty\Bigl[\left(f^{sc,1}_m\cos m\theta_{\b{p}\b{p}'}+f^{sc,2}_m\sin m\theta_{\b{p}\b{p}'}\right)\hat{p}_i'
+\left(f^{sc,3}_m\cos m\theta_{\b{p}\b{p}'}+f^{sc,4}_m\sin m\theta_{\b{p}\b{p}'}\right)\epsilon_{ji}\hat{p}_j'\Bigr],\\
V_{ij}(\hat{\b{p}},\hat{\b{p}}')&=\frac{1}{4}\sum_{m=0}^\infty\Bigl[f^{ss,1}_m\cos m\theta_{\b{p}\b{p}'}\delta_{ij}+f^{ss,3}_m\sin m\theta_{\b{p}\b{p}'}\epsilon_{ij}\nonumber\\
&+f^{ss,4}_m\cos m\theta_{\b{p}\b{p}'}\left(\epsilon_{kj}\hat{p}_i\hat{p}_k'
+\epsilon_{ki}\hat{p}_k\hat{p}_j'\right)
+f^{ss,5}_m\cos m\theta_{\b{p}\b{p}'}\left(\hat{p}_i\hat{p}_j'
-\epsilon_{ki}\epsilon_{lj}\hat{p}_k\hat{p}_l'\right)\Bigr].
\end{align}
Substituting these expressions into Eq.~(\ref{VbarSum}), we find
\begin{align}
\bar{V}(\hat{\b{p}},\hat{\b{p}}')=\sum_{l=0}^\infty\bar{f}_l\cos l\theta_{\b{p}\b{p}'},
\end{align}
where the projected Landau parameters $\bar{f}_l$ are given by Eq.~(9) in the main text.

\section{Equilibrium properties of helical Fermi liquids}
\label{sec:eqprop}

This section presents a detailed derivation of the physical properties of helical Fermi liquids from the projected Landau functional [Eq.~(8) of the main text]. Rather than as the coefficients of terms in a second-quantized interaction Hamiltonian operator, we would really like to think of $\bar{f}_l$ as the coefficients in this functional,
\begin{align}
\delta\bar{E}[\delta\bar{n}_\b{p}]=\int\frac{d^2p}{(2\pi)^2}(\epsilon^0_\b{p}-\mu)\delta\bar{n}_\b{p}+\frac{1}{2}\sum_{l=0}^\infty\int\frac{d^2p}{(2\pi)^2}\int\frac{d^2p'}{(2\pi)^2}\bar{f}_l\cos l\theta_{\b{p}\b{p}'}\delta\bar{n}_\b{p}\delta\bar{n}_{\b{p}'},
\end{align}
where $\epsilon^0_\b{p}=v_Fp$ and we have explicitly added a chemical potential term $\mu$. The (renormalized) quasiparticle energy $\epsilon_\b{p}$ is given by the functional derivative of the Landau functional with respect to the quasiparticle distribution function,
\begin{equation}
\epsilon_\b{p}=\frac{\delta\bar{E}}{\delta\bar{n}_\b{p}}
=\epsilon^0_\b{p}+\sum_{l=0}^\infty\int\frac{d^2p}{(2\pi)^2}\bar{f}_l\cos l\theta_{\b{p}\b{p}'}\delta\bar{n}_{\b{p}'}.
\label{QuasiParticleEnergy}
\end{equation}
It is important to note that $v_F$ is the renormalized Fermi velocity and not the bare (noninteracting) one, which is denoted by $v_F^0$.

\subsection{Specific heat}

We first investigate the specific heat. The derivation we use closely follows the standard derivation for the specific heat in standard Fermi liquids.\cite{PinesNozieres} The entropy density, $s$, is given by
\begin{equation}
s=-k_B\int\frac{d^2p}{(2\pi)^2}\big(\bar{n}_\b{p}\ln(\bar{n}_\b{p})-(1-\bar{n}_\b{p})\ln(1-\bar{n}_\b{p})\big),
\end{equation}
where $k_B$ is Boltzmann's constant. The variation is the entropy density is given by
\begin{equation}
\delta s=-k_B\int\frac{d^2p}{(2\pi)^2}\delta \bar{n}_{\b{p}}\ln\bigg(\frac{\bar{n}_\b{p}}{1-\bar{n}_\b{p}}\bigg)=-\frac{1}{T}\int\frac{d^2p}{(2\pi)^2}\delta \bar{n}_{\b{p}}(\epsilon_\b{p}-\mu),
\end{equation}
where $T$ is the temperature. The variation in particle density can be written as
\begin{equation}
\delta\bar{n}_\b{p}=\frac{\partial \bar{n}_\b{p}}{\partial \epsilon_\b{p}}\bigg(-\frac{\epsilon_\b{p}-\delta\mu}{T}\delta T+\delta\epsilon_\b{p}-\mu\bigg).
\label{density_var}
\end{equation}
To lowest order in $T$, we have 
\begin{equation}
\delta s =-\frac{1}{T^2}\int\frac{d^2p}{(2\pi)^2}\frac{\partial \bar{n}_\b{p}}{\partial \epsilon_\b{p}}(\epsilon_\b{p}-\mu)^2\delta T=-k_B^2g(\mu)\int_{-\infty}^{\infty}d\epsilon\bigg(\frac{\epsilon-\mu}{k_BT}\bigg)^2\frac{\partial \bar{n}}{\partial \epsilon}=\frac{\pi^2}{3}\rho(\mu)k_B^2T,
\end{equation}
where $\rho(\mu)$ is the density of states at the Fermi surface which is given by
\begin{equation}
\rho(\epsilon)=\int\frac{d^2p}{(2\pi)^2}\delta(\epsilon-\epsilon_\b{p}).
\end{equation}
For the noninteracting case we have $\rho(\mu)=\mu/2\pi v_F^2$. The specific heat is then
\begin{equation}
c_v=\frac{\pi^2}{3}\rho(\mu)k_B^2T.
\end{equation}
One then defines the electronic specific heat coefficient $\gamma$ as the zero-temperature limit of $c_v/T$,
\begin{align}
\gamma=\frac{1}{3}\pi^2k_B^2\rho(\epsilon_F).
\end{align}

\subsection{Compressibility}

We now turn to the electronic compressibility. Again, this derivation closely follows that for standard Fermi liquids. The compressibility $\kappa$ at zero temperature is defined as
\begin{equation}
\kappa=\frac{1}{n^2}\frac{\partial n}{\partial \mu},
\end{equation}
where $n$ is the density of electrons, given by $n=\int\frac{d^2p}{(2\pi)^2}\rho_\b{p}$ where $\rho_\b{p}=\sum_\sigma\langle c_{\b{p}\sigma}^\dag c_{\b{p}\sigma}^{\phantom\dagger}\rangle$. Projecting the field operators to the Fermi surface, we obtain $\rho_\b{p}=\bar{n}_\b{p}$ as expected, thus $\delta\rho_\b{p}=\delta\bar{n}_\b{p}$. At zero temperature, the density variation [Eq.~\eqref{density_var}] is
\begin{equation}
\delta\bar{n}_\b{p}=\frac{\partial \bar{n}_\b{p}}{\partial \epsilon_\b{p}}(\delta\epsilon_\b{p}-\delta\mu).
\label{var_2}
\end{equation}
The quantity $\frac{\partial \bar{n}_\b{p}}{\partial \epsilon_\b{p}}$ vanishes everywhere except at the Fermi surface and the variation of $\mu$ produces a variation of $\delta\bar{n}_\b{p}$ that is isotropic. Integrating Eq.~\eqref{var_2} over momentum, we find
\begin{equation}
\delta\bar{n}=\int\frac{d^2p}{(2\pi)^2}\delta\bar{n}_\b{p}=\int\frac{d^2p}{(2\pi)^2}\frac{\partial \bar{n}_\b{p}}{\partial \epsilon_\b{p}}(\delta\epsilon_\b{p}-\delta\mu)=\int\frac{d^2p}{(2\pi)^2}\frac{\partial \bar{n}_\b{p}}{\partial \epsilon_\b{p}}\left(\sum_{l=0}^\infty\int\frac{d^2p'}{(2\pi)^2}\bar{f}_l\cos l\theta_{\b{p}\b{p}'}\delta\bar{n}_{\b{p}'}-\delta\mu\right).
\end{equation}
For the last step, we used Eq.~\eqref{QuasiParticleEnergy} for $\delta\epsilon_\b{p}$. After integration over $\b{p}'$, only the $l=0$ contribution remains and we have
\begin{equation}
\int\frac{d^2p}{(2\pi)^2}\frac{\partial\bar{n}_\b{p}}{\partial \epsilon_\b{p}}\left(\sum_{l=0}^\infty\int\frac{d^2p'}{(2\pi)^2}\bar{f}_l\cos l\theta_{\b{p}\b{p}'}\delta\bar{n}_{\b{p}'}-\delta\mu\right)=\int\frac{d^2p}{(2\pi)^2}\frac{\partial \bar{n}_\b{p}}{\partial \epsilon_\b{p}}(\bar{f}_0\delta\bar{n}-\delta\mu).
\end{equation}
Defining dimensionless projected Landau parameters $\bar{F}_l$ as
\begin{equation}
\bar{F}_l\equiv \rho(\epsilon_F)\bar{f}_l\int_0^{2\pi}\frac{d\theta}{2\pi}\cos^2l\theta
=\left\{
\begin{array}{cc}
\rho(\epsilon_F)\bar{f}_0,&l=0, \\
\half\rho(\epsilon_F)\bar{f}_l,&l=1,2,3,\ldots
\end{array}
\right.,
\end{equation}
we arrive at
\begin{equation}
\kappa=\frac{\rho(\epsilon_F)}{n^2}\frac{1}{1+\bar{F}_0}.
\label{kappa}
\end{equation}

\subsection{Spin susceptibility}

We now investigate the spin susceptibility of a helical Fermi liquid. As mentioned in the main text, contrary to a standard Fermi liquid here the spin susceptibility is not strictly speaking a Fermi surface property. Therefore the present projected Fermi liquid theory can only correctly describe the spin susceptibility of the helical Fermi liquid in a certain limit to be seen below.

The total spin density of the helical Fermi liquid is $\b{s}=\int\frac{d^2p}{(2\pi)^2}\b{s}_\b{p}$ where
\begin{align}
\b{s}_\b{p}=\frac{1}{2}\langle c_\b{p}^\dag\bsigma c_\b{p}^{\phantom\dagger}\rangle.
\end{align}
Projecting the fermion operators to the Fermi surface, we obtain
\begin{align}
s^i_\b{p}&=\frac{1}{2}\epsilon_{ij}\hat{p}_j\bar{n}_\b{p},\hspace{5mm}i=1,2,\\
s^z_\b{p}&=0.
\end{align}
Therefore, we have $\delta s^i_\b{p}=\frac{1}{2}\epsilon_{ij}\hat{p}_j\delta\bar{n}_\b{p}$ and $\delta s^z_\b{p}=0$. As a result, our projected Fermi liquid theory will predict a zero out-of-plane susceptibility $\chi_{zz}=0$. For the in-plane susceptibility, consider applying an in-plane Zeeman term,
\begin{align}
\delta H=-\int\frac{d^2p}{(2\pi)^2}g\mu_B\b{B}\cdot\frac{1}{2}c_\b{p}^\dag\bsigma c_\b{p}^{\phantom\dagger},
\end{align}
where $\mu_B$ is the Bohr magneton and $g$ is the $g$-factor of the helical Fermi liquid. Ignoring constant terms, this leads to a change in the energy,
\begin{align}
\delta\bar{E}=\langle\delta H\rangle=-\int\frac{d^2p}{(2\pi)^2}g\mu_B\b{B}\cdot\delta\b{s}_\b{p}=\int\frac{d^2p}{(2\pi)^2}\delta\epsilon_\b{p}(\b{B})\delta\bar{n}_\b{p},
\end{align}
where the change in quasiparticle energy is
\begin{align}
\delta\epsilon_\b{p}(\b{B})=\epsilon_\b{p}(\b{B})-\epsilon_\b{p}=-\frac{1}{2}g\mu_B\b{B}\times\hat{\b{p}}.
\end{align}
The variation of $\delta\bar{n}_\b{p}$ is given by
\begin{equation}
\delta\bar{n}_\b{p}=\frac{\partial \bar{n}_{\b{p}}}{\partial \epsilon_\b{p}}(\delta \epsilon_\b{p}-\delta\mu),
\label{var_mag}
\end{equation}
where
\begin{equation}
\delta\epsilon_{\b{p}}=-\frac{1}{2}g\mu_B B_i\epsilon_{ij}\hat{p}_j+\sum_{l=0}^\infty\int\frac{d^2p'}{(2\pi)^2}
\bar{f}_l\cos l \theta_{\b{p}\b{p}'}\delta \bar{n}_{\b{p}'}.
\end{equation}
Since the chemical potential $\mu$ is a scalar and does not depend on the direction of the magnetic field, its variation can be ignored when calculating the linear susceptibility. We now introduce a renormalized $g$-factor $g_i(\b{p})$ that depends on quasiparticle momentum
\begin{equation}
\delta\epsilon_{\b{p}}=-\frac{1}{2}g_i(\b{p})\mu_B B_i=-\frac{1}{2}g\mu_B B_i\epsilon_{ij}\hat{p}_j+\sum_{l=0}^\infty\int\frac{d^2p'}{(2\pi)^2}\bar{f}_l\cos l \theta_{\b{p}\b{p}'}\delta \bar{n}_{\b{p}'}
\label{gen_g_fac}.
\end{equation}
Inserting Eq.~\eqref{var_mag} into Eq.~\eqref{gen_g_fac}, we find an integral equation for $g_i(\b{p})$,
\begin{equation}
\frac{g_i(\b{p})}{g}=\epsilon_{ij}\hat{p}_j+\sum_{l=0}^\infty\int\frac{d^2p'}{(2\pi)^2}\bar{f}_l\cos l \theta_{\b{p}\b{p}'}\frac{\partial \bar{n}_{\b{p}'}}{\partial \epsilon_{\b{p}'}}\frac{g_i(\b{p}')}{g}.
\label{int_for_g}
\end{equation}
We note that integral equations also appear for a partially spin-polarized Fermi liquid.\cite{Meyerovich92} The spin susceptibility is then
\begin{equation}
\chi_{ii}=\lim_{B\to 0}\frac{g\mu_B}{B}\int\frac{d^2p}{(2\pi)^2}\delta s^i_\b{p}=\frac{g^2\mu_B^2}{4}\int\frac{d^2p}{(2\pi)^2}\epsilon_{ij}\hat{p}_j\frac{\partial \bar{n}_{\b{p}}}{\partial \epsilon_\b{p}}\left(\epsilon_{ik}\hat{p}_k+\sum_{l=0}^\infty\int\frac{d^2p'}{(2\pi)^2}\bar{f}_l\cos l \theta_{\b{p}\b{p}'}\frac{\partial \bar{n}_{\b{p}'}}{\partial \epsilon_{\b{p}'}}g_i(\b{p}')\right).
\label{spin_sus}
\end{equation}
To make some progess in the interacting case, we assume a solution of the form
\begin{equation}
g(\b{p})_i=g_\textrm{eff}\epsilon_{ik}\hat{p}_k.
\end{equation}
Only the $l=1$ term will survive. The $l=1$ cosine term can be written as $\cos\theta_{\b{p}\b{p}'}=\hat{p}_x\hat{p}_x'+\hat{p}_y\hat{p}_y'$. After evaluating the angular integral in Eq.~\eqref{int_for_g}, we find
\begin{equation}
g_\textrm{eff}=\frac{g}{1+\bar{F}_1}.
\end{equation}
Turning to the spin susceptibility given by Eq.~\eqref{spin_sus}, we find
\begin{equation}
\chi_{ii}=\frac{g^2\mu_B^2}{4(1+\bar{F}_1)}\int\frac{d^2p}{(2\pi)^2}\epsilon_{ij}\hat{p}_j\frac{\partial \bar{n}_{\b{p}}}{\partial \epsilon_\b{p}}(\epsilon_{ik}\hat{p}_k)=\frac{1}{8}g^2\mu_B^2\rho(\epsilon_F)\frac{1}{1+\bar{F}_1},
\label{final_spin_sus}
\end{equation}
at zero temperature. We also find that $\chi_{ij}$ vanishes for $i\neq j$, which can be explicity seen from Eq.~\eqref{spin_sus}. This can be compared to the spin susceptibility of the noninteracting helical Fermi gas, which is derived using both helicities in Sec.~\ref{sec:spinfreedirac}.

\subsection{Pomeranchuk instabilities}

In this section we investigate the stability of the Fermi surface. The distortion of the Fermi surface can be characterized by an angular dependent Fermi wavevector,\cite{pomeranchuk1958}
\begin{equation}
p_F(\theta)-p_F=\sum_{l=-\infty}^\infty A_l e^{il\theta}.
\end{equation}
The change in energy is then
\begin{equation}
\delta\bar{E}[\delta\bar{n}_\b{p}]=\frac{\epsilon_F}{2\pi}\sum_{l=0}^{\infty}(1+\bar{F}_l)|A_l|^2,
\end{equation}
Here we have used that fact $A_{l}^{*}=A_{-l}$ since $p_F(\theta)$ is real. The Fermi surface is stable against spontaneous distortions only if $\delta\bar{E}>0$, i.e., if $\bar{F}_l>-1$ for all $l$. Let us discuss briefly the special case of the $l=2$ Pomeranchuk instability, which corresponds to a nematic instability.\cite{oganesyan2001} The $l=2$ projected interaction is of the form
\begin{align}
\delta\bar{E}=\frac{\bar{f}_2}{2}\int\frac{d^2p}{(2\pi)^2}\int\frac{d^2p'}{(2\pi)^2}\cos 2\theta_{\b{p}\b{p}'}\delta\bar{n}_\b{p}\delta\bar{n}_{\b{p}'}=\frac{\bar{f}_2}{4}\int\frac{d^2p}{(2\pi)^2}\int\frac{d^2p'}{(2\pi)^2}\tr\bar{Q}(\b{p})\bar{Q}(\b{p}'),
\end{align}
where
\begin{align}\label{ProjQij}
\bar{Q}_{ij}(\b{p})=(2\hat{p}_i\hat{p}_j-\delta_{ij})\delta\bar{n}_\b{p},
\end{align}
from which we can construct a traceless, symmetric 2D nematic order parameter\cite{oganesyan2001} $\bar{Q}_{ij}=\int\frac{d^2p}{(2\pi)^2}\bar{Q}_{ij}(\b{p})$. Interestingly, Eq.~(9) in the main text shows that this type of interaction can be obtained from $p$-wave unprojected spin-spin interactions, i.e., the unprojected Landau parameters $f^{ss,1}_1$ and $f^{ss,3}_1$. In fact, as mentioned in the main text [see Eq.~(19)] one can construct a 2D nematic order parameter in terms of the unprojected spin degrees of freedom,
\begin{align}
Q_{ij}(\b{p})=\hat{p}_i s^j_\b{p}+\hat{p}_j \delta s^i_\b{p}-\delta_{ij}\hat{\b{p}}\cdot\delta\b{s}_\b{p}.
\end{align}
This type of nematic order parameter was first considered in Ref.~\onlinecite{park2015} as a possible instability of 2D Majorana fermions, and its 3D analog was considered in Ref.~\onlinecite{fu2015} in the context of spin-orbit coupled 3D metals. One can then show that
\begin{align}
\frac{1}{2}\int\frac{d^2p}{(2\pi)^2}\int\frac{d^2p'}{(2\pi)^2}\tr Q(\b{p})Q(\b{p}')=\int\frac{d^2p}{(2\pi)^2}\int\frac{d^2p'}{(2\pi)^2}\left(\cos\theta_{\b{p}\b{p}'}\delta\b{s}_\b{p}\cdot\delta\b{s}_{\b{p}'}
-\sin\theta_{\b{p}\b{p}'}\delta\b{s}_\b{p}\times\delta\b{s}_{\b{p}'}\right),
\end{align}
which, comparing with Eq.~(\ref{VssFINAL}), corresponds to a spin-spin interaction with $f^{ss,1}_1=-f^{ss,3}_1\neq 0$. Because of Eq.~(9) in the main text, this corresponds indeed to a nonzero contribution to $\bar{f}_2$. In fact, if we project $Q_{ij}(\b{p})$ to the Fermi surface in the sense of replacing $\delta s^i_\b{p}$ by its expectation value on the Fermi surface $\langle \delta s^i_\b{p}\rangle=\frac{1}{2}\epsilon_{ij}\hat{p}_j$, we obtain
\begin{align}
\langle Q_{ij}(\b{p})\rangle=\frac{1}{2}\left(\begin{array}{cc}
2\hat{p}_x\hat{p}_y & \hat{p}_y^2-\hat{p}_x^2 \\
\hat{p}_y^2-\hat{p}_x^2 & -2\hat{p}_x\hat{p}_y
\end{array}\right),
\end{align}
which is essentially equivalent to Eq.~(\ref{ProjQij}) except for a rotation by $\pi/4$ about the $z$ axis: by rotating $\hat{\b{p}}\rightarrow R_{\pi/4}\hat{\b{p}}$, we have $\langle Q_{ij}(\b{p})\rangle\rightarrow\frac{1}{2}(2\hat{p}_i\hat{p}_j-\delta_{ij})$.

\subsection{Renormalized velocity}\label{sec:renvel}

In this section we consider the renormalization of the Fermi velocity. We begin by considering a microscopic Hamiltonian in first quantization,
\begin{equation}
H=v_F^0\hat{\b{z}}\cdot(\b{\sigma}\times\b{p})+H_{\mathrm{int}},
\end{equation}
where the electron-electron interaction $H_{\mathrm{int}}$ is assumed to not depend on momentum. The renormalization of the Fermi velocity is similar, in spirit, to the renormalization of the quasiparticle mass in normal Fermi liquids. The derivation, however, is quite different because spin-orbit coupling breaks Galilean invariance. Following Ref.~\onlinecite{ashrafi2013}, we use the fact that the total flux of quasiparticles is equal to the total flux of particles. To find the velocity operator of the particles, we use the commutation relation
\begin{equation}
\b{v}_e=-\frac{i}{\hbar}[\b{x},H]= v_F^0 (\hat{\b{z}}\times\bsigma).
\end{equation}
Because the interaction is momentum-independent, we have $[\b{x},H_\textrm{int}]=0$ and the velocity operator is the same as in the absence of interactions. By equating the total flux of particles and quasiparticles we find
\begin{equation}
\int\frac{d^2p}{(2\pi)^2}v_F^0 (\hat{\b{z}}\times\langle\psi_\b{p}^\dag\bsigma\psi_\b{p}^{\phantom\dagger}\rangle)=\int\frac{d^2p}{(2\pi)^2}\bar{n}_\b{p}\nabla_\b{p}\epsilon_\b{p}
\label{p_flux}.
\end{equation}
After projecting the fermion operators to the Fermi surface and varying both sides of Eq.~\eqref{p_flux}, we find
\begin{equation}
\int\frac{d^2p}{(2\pi)^2} v^0_F \hat{\b{p}}\delta \bar{n}_\b{p}=\int\frac{d^2p}{(2\pi)^2}(\nabla_\b{p}\epsilon^0_\b{p})\delta\bar{n}_\b{p}-\sum_{l=0}^{\infty}
\int\frac{d^2p}{(2\pi)^2}\int\frac{d^2p'}{(2\pi)^2}\bar{f}_l\cos l\theta_{\b{p}\b{p}'}(\nabla_\b{p}\bar{n}^0_\b{p})\delta\bar{n}_\b{p}.
\end{equation}
After relabeling $\b{p}\rightarrow\b{p}'$ and equating the integrands, since the variation of $\bar{n}_\b{p}$ is arbitrary, we find
\begin{equation}
 v^0_F \hat{\b{p}}=\nabla_\b{p}\epsilon^0_\b{p}-\sum_{l=0}^{\infty}\int\frac{d^2p'}{(2\pi)^2}\bar{f}_l\cos l\theta_{\b{p}\b{p}'}\nabla_{\b{p}'}\bar{n}^0_{\b{p}'}
 =v_F\left(\hat{\b{p}}+\sum_{l=0}^{\infty}\int\frac{d^2p'}{(2\pi)^2}\bar{f}_l\cos l\theta_{\b{p}\b{p}'}\delta(v_Fp-\mu)\hat{\b{p}}'\right).
\end{equation}
Multiplying both sides by $\hat{\b{p}}$, we obtain
\begin{equation}
v_F^0 =v_F\left(1+\sum_{l=0}^{\infty}\int\frac{d^2p'}{(2\pi)^2}\bar{f}_l\cos l\theta_{\b{p}\b{p}'}\delta(v_Fp'-\mu)\cos\theta_{\b{p}\b{p}'}\right)
=v_F(1+\bar{F}_1).
\end{equation}
Only the $l=1$ term contributes, and we arrive at Eq.~(14) of the main text.

\section{Spin susceptibility of the noninteracting helical Fermi gas}
\label{sec:spinfreedirac}

In this section we calculate the spin susceptibility of the noninteracting helical Fermi gas while taking both helicities into account (i.e., without projecting out the negative helicity part).

\subsection{Out-of-plane spin susceptibility}

We first consider the out-of-plane susceptibility at zero temperature. We consider a free Dirac system with a Zeeman term, ignoring orbital effects of the magnetic field. (The combined effects of Zeeman and orbital couplings on the spin susceptibility were studied in Ref.~\onlinecite{vazifeh2012}.) Landau quantization is expected to dominate only at very low fields. Spefically the orbital contribution will dominate if
\begin{equation}
\hbar v_F \sqrt{\frac{eB}{\hbar}}>g\mu_BB.
\end{equation}
Experimental parameters for $\mathrm{Bi}_2\mathrm{Se}_3$, $\mathrm{Sb}_2\mathrm{Te}_3$, and $\mathrm{Bi}_2\mathrm{Te}_3$ exposed to ambient conditions put this scale on the order of $10^{-4}$~T, thus for those systems, the Zeeman effect will dominate under typical experimental conditions.\cite{zhao2014} (When exposed to ambient conditions the Fermi velocity can decrease by two orders of magnitude,\cite{PhysRevLett.109.247604} which allows the Zeeman term to dominate down to very small fields.) The Hamiltonian of the system is given by
\begin{equation}
H=\int\frac{d^2k}{(2\pi)^2}c_\b{k}^\dag \left(h(\b{k})-\mu-\half g\mu_B B\sigma_z\right)c_\b{k}^{\phantom\dagger},
\end{equation}
where $c_\b{k}=(c_{\b{k}\ua},c_{\b{k}\da})$ is a two-component Dirac spinor, and
\begin{align}
h(\b{k})=v_F^0\hat{\b{z}}\cdot(\b{\sigma}\times\b{k})=v_F^0\left(
\begin{array}{cc}
0 & ike^{-i\theta_\b{k}} \\
-ike^{i\theta_\b{k}} & 0
\end{array}\right).
\end{align}
We note that the out-of-plane Zeeman term is not diagonal in the helicity basis and thus cannot be captured by our theory. This Hamiltonian can be diagonalized exactly. The full Hamiltonian can be written as
\begin{align}
H=v_F^0\int\frac{d^2k}{(2\pi)^2}c_\b{k}^\dag\left(
\begin{array}{cc}
-\frac{g\mu_BB}{2 v_F^0} & ike^{-i\theta_\b{k}} \\
-ike^{i\theta_\b{k}} & \frac{g\mu_BB}{2 v_F^0} 
\end{array}\right)c_\b{k}^{\phantom\dagger}.
\end{align}
The eigenenergies $E_\chi(\b{k})=\chi v_F^0k+\mathcal{O}(B^2)$ do not change to linear order in out-of-plane field strength. The eigenvectors for a given chirality $\chi=\pm 1$ are
\begin{align}
|\psi_\chi(\b{k})\rangle=\frac{1}{\sqrt{2}}\left(
\begin{array}{cc}
i\chi\left(1-\chi\frac{g\mu_BB}{2 v_F^0k}\right)e^{-i\theta_{\b{k}}} \\
1
\end{array}\right).
\end{align}
We now calculate the expectation value of the Pauli matrices for a given chirality to lowest order in field strength. This expectation value is proportional to the magnetization. We find
\begin{align}
\langle\psi_\chi(\b{k})|\sigma_z|\psi_\chi(\b{k})\rangle&=-\chi\frac{g\mu_BB}{2 v_F^0k}, \nonumber \\
\langle\psi_\chi(\b{k})|\sigma_x|\psi_\chi(\b{k})\rangle&=\chi\left(1+\chi\frac{g\mu_BB}{2 v_F^0k}\right)\sin\theta_\b{k}, \nonumber \\
\langle\psi_\chi(\b{k})|\sigma_y|\psi_\chi(\b{k})\rangle&=\chi\left(1+\chi\frac{g\mu_BB}{2 v_F^0k}\right)\cos\theta_\b{k}.
\end{align}
Summing over momentum and chirality, we find 
\begin{equation}
\langle\sigma_z\rangle=-\sum_{\chi}\int\frac{d^2k}{(2\pi)^2}\langle\psi_\chi(\b{k})|\sigma_z|\psi_\chi(\b{k})\rangle n_F(E_\chi(\b{k}))=\int_0^{\Lambda/v_F^0}\frac{dk}{2\pi}\frac{g\mu_BB}{2 v_F^0}\left[\Theta(\epsilon_F-v_F^0 k)-\Theta(\epsilon_F+v_F^0 k)\right],
\end{equation}
where $n_F(\epsilon)=(e^{\beta(\epsilon-\mu)}+1)^{-1}$ is the Fermi function (evaluated at zero temperature $\beta\rightarrow\infty$), and $\langle\sigma_x\rangle=\langle\sigma_y\rangle=0$ due to the angular integral vanishing. We have also introduced a high-energy cutoff $\Lambda$. Evaluating the integrals, we find
\begin{equation}
\langle\sigma_z\rangle=\frac{g\mu_BB}{4\pi (v_F^0)^2}(\Lambda-\epsilon_F)=\frac{g\mu_B B}{2}[\rho(\Lambda)-\rho(\epsilon_F)],
\end{equation}
where $\rho(\epsilon)=|\epsilon|/2\pi(v_F^0)^2$ is the density of states of the helical Fermi gas. This corresponds to an out-of-plane susceptibility
\begin{align}
\chi_{zz}=\frac{\half g\mu_B\langle\sigma_z\rangle}{B}=\frac{1}{4}g^2\mu_B^2[\rho(\Lambda)-\rho(\epsilon_F)].
\end{align}
In the limit that $\epsilon_F\rightarrow\Lambda$, the out-of-plane spin susceptibility vanishes in agreement with the projected helical Fermi theory.

\subsection{In-plane spin susceptibility}

In this section we consider the in-plane susceptibility. We take the in-plane magnetic field to be in the $x$-direction without loss of generality due to $SO(2)$ rotation symmetry. The Hamiltonian we consider is
\begin{equation}
H=\int\frac{d^2k}{(2\pi)^2}c_\b{k}^\dag \left(h(\b{k})-\mu-\half g\mu_B B\sigma_x\right)c_\b{k}^{\phantom\dagger}.
\end{equation}
The energy of an eigenstate of given chirality to linear order in field strength is
\begin{equation}
E_{\chi}(\b{k})=\chi v_F^0 k-\chi\sin\theta_\b{k}\frac{g\mu_BB}{2}+\mathcal{O}(B^2),
\label{in-plane-energy}
\end{equation}
and the eigenstate of a given chirality is
\begin{align}
|\psi_\chi(\b{k})\rangle=
\frac{1}{\sqrt{2}}\left(
\begin{array}{cc}
\chi\left(i-\cos\theta_\b{k}\frac{g\mu_BB}{2 v_F^0k}\right)e^{-i\theta_{\b{k}}} \\
1
\end{array}\right)+\mathcal{O}(B^2).
\end{align}
We now calculate the expectation values of the Pauli matrices. We find to linear order in field strength
\begin{align}
\langle\psi_\chi(\b{k})|\sigma_z|\psi_\chi(\b{k})\rangle&=0,\nonumber \\
\langle\psi_\chi(\b{k})|\sigma_x|\psi_\chi(\b{k})\rangle&=\chi\bigg(\sin\theta_\b{k}-\cos^2\theta_\b{k}
\frac{g\mu_BB}{2 v_F^0 k}\bigg), \nonumber \\
\langle\psi_\chi(\b{k})|\sigma_y|\psi_\chi(\b{k})\rangle&=\chi\bigg(\cos\theta_\b{k}-\cos\theta_\b{k}
\sin\theta_\b{k}\frac{g\mu_BB}{2 v_F^0 k}\bigg).
\end{align}
Summing over momentum and chirality we find
\begin{equation}
\langle\sigma_x\rangle=\int\frac{d^2k}{(2\pi)^2}\bigg(\sin\theta_\b{k}-\cos\theta_\b{k}^2\frac{g\mu_BB}{2 v_F^0 k}\bigg) \left[\Theta\left(\epsilon_F- v_F^0 k+\sin\theta_\b{k}\frac{g\mu_BB}{2}\right)-\Theta\left(\epsilon_F+v_F^0 k-\sin\theta_\b{k}\frac{g\mu_BB}{2}\right)\right].
\end{equation}
Simplifying the equation by using the fact that $\Theta(\epsilon_F+ v_F^0 k-\sin\theta_\b{k}\frac{g\mu_BB}{2})$ is always one for a large Fermi energy $\epsilon_F$, we find
\begin{equation}
\langle\sigma_x\rangle=\int \frac{d\theta}{2\pi}\left(\int_0^{k_F+\sin\theta_\b{k}\frac{g\mu_BB}{2 v_F^0}}\frac{dk}{2\pi}-\int_0^{\frac{\Lambda}{ v_F^0}}\frac{dk}{2\pi}\right)\bigg(\sin\theta_\b{k}k-\cos^2\theta_\b{k}\frac{g\mu_BB}{2 v_F^0}\bigg).
\end{equation}
Performing the integration, we find
\begin{equation}
\langle\sigma_x\rangle=\frac{g\mu_B B}{4}\rho(\Lambda),
\end{equation}
which gives for the susceptibility
\begin{equation}
\chi_{xx}=\frac{1}{8}g^2\mu_B^2 \rho(\Lambda).
\end{equation}
This agrees with the result obtained from our projected helical Fermi liquid theory [Eq.~\eqref{final_spin_sus} with $\bar{F}_1=0$] when $\epsilon_F\rightarrow\Lambda$.

\section{Collective modes in a helical Fermi liquid}
\label{Coll_MODES}

In this section we investigate the collective modes of a helical Fermi liquid in the presence of a monochromatic external scalar potential,  $U(\b{r},t)=U e^{i(\b{q}\cdot\b{r}-\omega t)}$ with $q\ll p_F$ and $\omega \ll \epsilon_F=v_Fp_F$. The quasiparticle distribution function $\bar{n}_{\b{p}}$ obeys the kinetic equation
\begin{equation}
\frac{\partial \bar{n}_\b{p}(\b{r},t)}{\partial t}+\nabla_{\b{p}}\epsilon_\b{p}(\b{r},t)\cdot\nabla_\b{r}\bar{n}_\b{p}(\b{r},t)-\nabla_{\b{r}}\epsilon_{\b{p}}(\b{r},t)\cdot\nabla_{\b{p}}\bar{n}_\b{p}(\b{r},t)=I[\bar{n}_\b{p}],
\label{Kin_eq}
\end{equation}
where $I[\bar{n}_\b{p}]$ is the collision integral. In the presence of an external scalar potential, $\bar{n}_{\b{p}}(\b{r},t)$ and the quasiparticle energy are given by
\begin{equation}
\bar{n}_{\b{p}}(\b{r},t)=\bar{n}^0_\b{p}+\delta\bar{n}_{\b{p}}(\b{r},t), ~~~~~\epsilon_\b{p}(\b{r},t)=\epsilon_{\b{p}}^0+U(\b{r},t)+\sum_{l=0}^\infty\int\frac{d^2p}{(2\pi)^2}\bar{f}_l\cos l\theta_{\b{p}\b{p}'}\delta\bar{n}_{\b{p}'}.
\end{equation}
In general, the collision integral involves scattering between states with different helicities, and one must keep both helicities. However, if the relaxation-time approximation is valid, scattering between states with different helicities can be neglected.

\subsection{First sound}

We first consider collective modes in the hydrodynamic regime (regular sound waves), i.e., $\omega\tau \ll 1$ where $\tau$ is the quasiparticle collision time. Our goal is to find the sound velocity in the presence of quasiparticle interactions. This derivation does not follow the standard derivation for regular Fermi liquids due to the lack of Galilean invariance. We first obtain the local momentum conservation law. To do so, we first multiply Eq.~\eqref{Kin_eq} by $p_i$ and then integrate over $\b{p}$, which gives
\begin{equation}
\frac{\partial g_i(\b{r},t)}{\partial t}+\frac{\partial T_{ij}(\b{r},t)}{\partial r_j}+\int\frac{d^2p}{(2\pi)^2}\frac{\partial\epsilon_{\b{p}}(\b{r},t)}{\partial r_i}\bar{n}_\b{p}(\b{r},t)=0,
\label{mom_con}
\end{equation}
where
\begin{equation}
g_i(\b{r},t)=\int\frac{d^2p}{(2\pi)^2}p_i\bar{n}_\b{p}(\b{r},t),~~~~T_{ij}(\b{r},t)=\int\frac{d^2p}{(2\pi)^2}p_i\frac{\partial\epsilon_\b{p}(\b{r},t)}{\partial p_j}\bar{n}_\b{p}(\b{r},t),
\end{equation}
where $g_i$ is the momentum density. The integral of the collision term vanishes due to conservation of quasiparticle momentum. We can rewrite this equation as
\begin{equation}
\frac{\partial g_i(\b{r},t)}{\partial t}+\frac{\partial\Pi_{ij}(\b{r},t)}{\partial r_j}+\bar{n}(\b{r},t)\frac{\partial U(\b{r},t)}{\partial r_i}=0,
\label{mom_con}
\end{equation}
where $\bar{n}(\b{r},t)=\int\frac{d^2p}{(2\pi)^2}\bar{n}_\b{p}(\b{r},t)$ and $\Pi_{ij}$ is the total stress tensor, given by
\begin{equation}
\Pi_{ij}=T_{ij}-\delta_{ij}\int\frac{d^2p}{(2\pi)^2}U(\b{r},t)\bar{n}_\b{p}(\b{r},t).
\end{equation}
In general, the stress tensor is of the form
\begin{equation}
\Pi_{ij}=P\delta_{ij}-\alpha_{ij},
\end{equation}
where $P$ is the pressure and $\alpha_{ij}$ is the dissipative part of the stress tensor. We neglect $\alpha$ for the rest of this work as it does not have an effect on the sound velocity. We now turn to the local energy conservation law. Multiplying Eq.~\eqref{Kin_eq} by $\epsilon_{\b{p}}(\b{r},t)$ and then integrating over $\b{p}$ gives
\begin{equation}
\int\frac{d^2p}{(2\pi)^2}\epsilon_{\b{p}}(\b{r},t)\frac{\partial \bar{n}_\b{p}(\b{r},t)}{\partial t}+\frac{\partial}{\partial r_j}\int\frac{d^2p}{(2\pi)^2}\epsilon_{\b{p}}(\b{r},t)\frac{\partial \epsilon_{\b{p}}(\b{r},t)}{\partial p_j}\bar{n}_\b{p}(\b{r},t)=0.
\label{eng_con}
\end{equation}
We now linearize Eq.~\eqref{mom_con} and \eqref{eng_con}. 
First we define,
\begin{equation}
\int_0^\infty\frac{dp\,p}{(2\pi)^2}\delta\bar{n}_\b{p}(\b{r},t)=\delta\bar{n}(\b{\hat{p}},\b{r},t),~~~~~\int_0^\infty\frac{dp\,p}{(2\pi)^2}p\delta\bar{n}_\b{p}(\b{r},t)=\delta\Omega(\b{\hat{p}},\b{r},t),
\label{mode_ex_n}
\end{equation}
and mode expand $\delta\bar{n}(\b{\hat{p}},\b{r},t)$ and $\delta\Omega(\b{\hat{p}},\b{r},t)$ as
\begin{equation}
\delta\bar{n}(\b{\hat{p}},\b{r},t)=\sum_{n=-\infty}^{\infty}B_n(\b{r},t)e^{in\theta},~~~~~\delta\Omega(\b{\hat{p}},\b{r},t)=\sum_{n=-\infty}^{\infty}C_n(\b{r},t)e^{in\theta}.
\end{equation}
The total density and energy fluctuations are then
\begin{equation}
\delta \bar{n}(\b{r},t)=2\pi B_0(\b{r},t),~~~~~\delta\epsilon(\b{r},t)=2\pi v_F C_0(\b{r},t).
\label{mode_den}
\end{equation}
We note that the energy and density fluctuations are related by the equation of state. To linear order, we have
\begin{equation}
\delta\epsilon(\b{r},t)=\frac{\partial f^0(n^0)}{\partial n^0}\delta \bar{n}(\b{r},t)=\mu \delta\bar{n}(\b{r},t),
\label{eq_of_state}
\end{equation}
where $f^0=\frac{4}{3}\pi^{1/2}n^{3/2}$ is the equation of state in the noninteracting limit. Upon linearizing Eq.~\eqref{mom_con} we obtain
\begin{equation}
\frac12\mu\partial_x \delta n(\b{r},t)(1+\bar{F}_0)+2\pi \partial_t(\mathrm{Re}\,C_{1}(\b{r},t))=-n^0\partial_xU,
\label{Mom_Lin_x}
\end{equation}
and
\begin{equation}
\frac12\mu\partial_y\delta n(\b{r},t)(1+\bar{F}_0)-2\pi  \partial_t (\mathrm{Im}\,C_{1}(\b{r},t))=-n^0\partial_yU.
\label{Mom_Lin_y}
\end{equation}
Here we have used $\delta P=\frac{n}{\partial n/\partial \mu}\delta n$, which is valid since the system is in local thermodynamic equilibrium, and
\begin{equation}
\partial_j \delta P(\b{r},t)\delta_{ij}=\frac12\mu\partial_i\delta n(\b{r},t)(1+\bar{F}_0),
\end{equation}
which follows from Eq.~(\ref{kappa}) for the compressibility. Linearizing Eq.~\eqref{eng_con} gives
\begin{align}
\partial_t \delta\epsilon(\b{r},t)+2\pi v_F^2\partial_x(\mathrm{Re}\,C_{1}(\b{r},t))(1+\bar{F}_1)-2\pi v_F^2\partial_y(\mathrm{Im}\,C_1(\b{r},t))(1+\bar{F}_1)=0.
\label{Lin_EN}
\end{align}
Taking the temporal derivative of Eq.~\eqref{Lin_EN}, the $x$-derivative of Eq.~\eqref{Mom_Lin_x} and the $y$-derivative of Eq.~\eqref{Mom_Lin_y}, and substituting Eq.~\eqref{Mom_Lin_x} and \eqref{Mom_Lin_y} into Eq.~\eqref{Lin_EN}, we obtain
\begin{align}
\partial_t^2\delta\epsilon(\b{r},t)-\frac12\mu v_F^2\nabla^2\delta n(\b{r},t)(1+\bar{F}_1)(1+\bar{F}_0)=v_F^2n^0\nabla^2U(1+\bar{F}_1).
\label{Lin_EN}
\end{align}
Using Eq.~\eqref{eq_of_state}, we find the equation of motion for the density fluctuations to be
\begin{align}
\partial_t^2n(\b{r},t)-\frac12 v_F^2\nabla^2\delta n(\b{r},t)(1+\bar{F}_1)(1+\bar{F}_0)=\frac{v_F^2}{\mu}n^0\nabla^2U(1+\bar{F}_1).
\label{Lin_EN}
\end{align}
Thus the velocity of first/hydrodynamic sound in the presence of quasiparticle interactions is given by
\begin{equation}
c_1=v_F\sqrt{\half(1+\bar{F}_0)(1+\bar{F}_1)}.
\end{equation}

\subsection{Zero sound}

We now turn to the collisionless regime $\omega\tau\gg 1$. Solving Eq.~\eqref{Kin_eq} by Fourier transform for $U=0$, we find
\begin{equation}
(\omega-\b{q}\cdot\b{v}_{\b{p}})\delta \bar{n}_\b{p}+\b{q}\cdot \b{v}_{\b{p}}\frac{\partial \bar{n}^0_{\b{p}}}{\partial \epsilon_{\b{p}}}\sum_{l=0}\int\frac{d^2p'}{(2\pi)^2} \bar{f}_l\cos\theta_{\b{p}\b{p}'}\delta\bar{n}_{\b{p}'}=0,
\end{equation}
where $\b{v}_\b{p}=v_F\hat{\b{p}}$. Following the standard approach in regular FLs, we assume a solution of $\delta\bar{n}_\b{p}$ of the form
\begin{equation}
\delta\bar{n}_\b{p}=-\frac{\partial \bar{n}^0_{\b{p}}}{\partial \epsilon_{\b{p}}}\nu_\b{p},
\end{equation}
and expand $\nu_\b{p}$ as
\begin{equation}
\nu_\b{p}=\sum_{m=-\infty}^{\infty}e^{im\theta}\bar{\nu}_m.
\end{equation}
We then have
\begin{equation}
(\omega-\b{q}\cdot\b{v}_{\b{p}})\nu_\b{p}-\b{q}\cdot \b{v}_{\b{p}}\sum_{l=0}^\infty\rho(\epsilon_F)\bar{f}_l
\int_0^{2\pi}\frac{d\theta_{\b{p}'}}{2\pi}
\cos l\theta_{\b{p}\b{p}'}\nu_{\b{p}'}=0.
\label{Gen_SOL}
\end{equation}
Choosing a system of coordinates for $\b{p}$ such that $\b{q}\cdot\hat{\b{p}}=q\cos\theta_\b{p}$, and defining the dimensionless variable $s=\omega/v_Fq$, we obtain
\begin{align}
\nu_\b{p}=\frac{\cos\theta_\b{p}}{s-\cos\theta_\b{p}}\sum_{l=0}^\infty\rho(\epsilon_F)\bar{f}_l
\int_0^{2\pi}\frac{d\theta_{\b{p}'}}{2\pi}
\cos l\theta_{\b{p}\b{p}'}\nu_{\b{p}'}=0.
\end{align}
We first consider the case that $\bar{F}_0$ only is nonzero. Integrating over both $\theta_\b{p}$ and $\theta_{\b{p}'}$, we obtain
the equation
\begin{align}
\left(1+\bar{F}_0\Omega_0(s)\right)\bar{\nu}_0=0,
\end{align}
where we define the dimensionless function
\begin{align}
\Omega_l(s)=\int_0^{2\pi}\frac{d\theta}{2\pi}\frac{\cos\theta\cos l\theta}{\cos\theta-s}=\Omega_{-l}(s),
\end{align}
which is easily evaluated for the first few values $l=0,1,2$,
\begin{align}
\Omega_0(s)&=1-\frac{s}{\sqrt{s^2-1}},\\
\Omega_1(s)&=s\Omega_0(s),\\
\Omega_2(s)&=1+(2s^2-1)\Omega_0(s).
\end{align}
A nontrivial solution $\bar{\nu}_0\neq 0$ requires
\begin{align}
1+\bar{F}_0\Omega(s)=0,
\end{align}
which is easily solved to give
\begin{align}
\frac{\omega}{v_Fq}=\frac{1+\bar{F}_0^{-1}}{\sqrt{(1+\bar{F}_0^{-1})^2-1}}\equiv\frac{c_0}{v_F},
\end{align}
where $c_0$ is the velocity of zero sound. A zero sound mode thus exists for all positive (repulsive) values of $\bar{F}_0$. Because $\Omega_0(s)$ is real only for $s>1$, for an undamped mode one must restrict oneself to $s>1$. Simple expressions can be obtained in the limits of strong and weak interaction,
\begin{align}
c_0&\approx v_F\sqrt{\frac{\bar{F}_0}{2}},\hspace{5mm}\bar{F}_0\rightarrow\infty,\\
c_0&\approx v_F\left(1+\frac{1}{2}\bar{F}_0^2\right),\hspace{5mm}\bar{F}_0\rightarrow 0.
\end{align}

We now consider turning on a nonzero value of $\bar{F}_1$ in addition to a positive $\bar{F}_0$. One then obtains three coupled equations for $\bar{\nu}_0,\bar{\nu}_1,\bar{\nu}_{-1}$,
\begin{align}
\left(1+\bar{F}_0\Omega_0(s)\right)\bar{\nu}_0
+\bar{F}_1\Omega_1(s)(\bar{\nu}_1+\bar{\nu}_{-1})&=0,\label{ZSeq1}\\
2\bar{F}_0\Omega_1(s)\bar{\nu}_0+\left(1+\bar{F}_1\Omega_0(s)
+\bar{F}_1\Omega_2\right)(\bar{\nu}_1+\bar{\nu}_{-1})&=0,\label{ZSeq2}\\
\left(1+\bar{F}_1\Omega_0(s)-\bar{F}_1\Omega_2(s)\right)
(\bar{\nu}_1-\bar{\nu}_{-1})&=0.
\end{align}
We see that the $l=0$ mode $\bar{\nu}_0$ and the symmetric combination of the $l=\pm 1$ modes $\bar{\nu}_1+\bar{\nu}_{-1}$ are coupled by the first two equations, while the antisymmetric combination $\bar{\nu}_1-\bar{\nu}_{-1}$ decouples. The equation for the latter would also be found in a model with a pure $\bar{F}_1$ interaction. Since we are primarily interested in the effects of a nonzero $\bar{F}_1$ interaction on the $l=0$ mode found earlier, we will focus on the first two equations (\ref{ZSeq1})-(\ref{ZSeq2}). The condition of a nontrivial solution for $\bar{\nu}_0$ and $\bar{\nu}_1+\bar{\nu}_{-1}$ gives
\begin{align}
\frac{1}{\bar{F}_0}=-\frac{(1+\bar{F}_1)\Omega_0(s)}{1+\bar{F}_1(1+2s^2\Omega_0(s))}.
\end{align}
Consider negative values of $\bar{F}_1$. One can show that $\Omega_0(s)$ and $1+2s^2\Omega_0(s)$ are negative for all $s>1$. For $\bar{F}_1<0$, we have
\begin{align}
\frac{1}{\bar{F}_0}=\frac{(1-|\bar{F}_1|)|\Omega_0(s)|}{1+|\bar{F}_1||1+2s^2\Omega_0(s)|}.
\end{align}
The right-hand side of this expression becomes negative for $\bar{F}_1<-1$, implying that zero sound is destroyed for sufficiently attractive values of $\bar{F}_1$. Given Eq.~(11) in the main text, this can occur, for instance, due to sufficiently attractive microscopic $l=0$ interactions in the spin channel, e.g., $f_0^{ss,1}$ sufficiently negative. The disappearance of zero sound due to sufficiently attractive interactions in the spin channel was also found in a microscopic study of the helical Fermi liquid.\cite{raghu2010}

\subsection{Determining $\bar{F}_1$ from first/zero sound}

As discussed in Sec.~\ref{sec:renvel}, there is an operator identity that relates the electron velocity operator $\b{v}_e$ to the electron spin $\bsigma$. This identity is valid in the presence of interactions, but only involves the noninteracting Fermi velocity $v_F^0$. Combined with the continuity equation $\partial_t n_\b{q}=-iqj_\b{q}^L$ where $n_\b{q}$ is the density operator and $j_\b{q}^L=\hat{\b{q}}\cdot\b{j}_\b{q}$ is the longitudinal current density operator, this yields the identity $\partial_tn_\b{q}=-iv_F^0qs_\b{q}^T$ where $s_\b{q}^T=\hat{\b{z}}\cdot(\bsigma\times\hat{\b{q}})$ is the transverse spin density operator. Passing to the frequency domain, this gives\cite{raghu2010}
\begin{align}\label{ratiosn}
\frac{\omega}{v_F^0q}=\frac{s_\b{q}^T}{n_\b{q}},
\end{align}
where the right-hand side is now interpreted as a ratio of expectation values. In Ref.~\onlinecite{raghu2010}, the authors suggest generating a spin-density wave of momentum $q$ and amplitude $s_\b{q}^T$ with a spin grating. In the presence of a collective mode of frequency $\omega=c_sq$ where $c_s$ is the sound velocity ($c_1$ or $c_0$ depending on whether one is in the hydrodynamic or collisionless regime), this will generate a long-lived density wave at momentum $q$ whose amplitude $n_\b{q}$ can in principle be measured. Using Eq.~(\ref{ratiosn}), the ratio of amplitudes of the original spin-density wave and induced density wave should be given by
\begin{align}
\frac{s_\b{q}^T}{n_\b{q}}=\frac{c_s}{v_F^0}=\frac{1}{1+\bar{F}_1}\frac{c_s}{v_F}.
\end{align}
Assuming for instance that one is in the hydrodynamic regime $\omega\tau\ll 1$, one would get
\begin{align}
\frac{s_\b{q}^T}{n_\b{q}}=\sqrt{\half\left(\frac{1+\bar{F}_0}{1+\bar{F}_1}\right)},
\end{align}
such that the value of $\bar{F}_1$ can be extracted from a measurement of the amplitude ratio, assuming that $\bar{F}_0$ is known from heat capacity and electronic compressibility measurements, as explained in the main text. While the sound modes give a $q$-independent ratio of amplitudes, the spin plasmon mode\cite{raghu2010} due to the unscreened Coulomb interaction gives a ratio $s_\b{q}^T/n_\b{q}\propto 1/\sqrt{q}$, which can in principle be used to discriminate between the two types of collective modes.

\bibliography{HFL}